\pdfoutput=1  
\documentclass[amstex,twocolumn,superscriptaddress,floats,floatfix,aps,pra, nofootinbib]{revtex4-2}
\usepackage{amssymb}
\usepackage{amsmath}
\usepackage{braket}
\usepackage{calc}
\usepackage{ragged2e}
\usepackage[justification=justified,singlelinecheck=false]{caption}
\usepackage{graphicx}
\usepackage{bm}
\usepackage{appendix}
\usepackage[dvipsnames]{xcolor}
\usepackage{titlesec}
\usepackage{subcaption}
\usepackage{soul}
\usepackage{xcolor}
\usepackage{pgfplots}
\pgfplotsset{compat=1.17}
\usepackage[most]{tcolorbox}
\usepackage{comment}
\sethlcolor{yellow}
\titleformat{\paragraph}
{\normalfont\normalsize}{\theparagraph}{1em}{}
\titlespacing*{\paragraph}
{0pt}{3.25ex plus 1ex minus .2ex}{1.5ex plus .2ex}

\def\be{
\begin{align} }
  \def\ee{
\end{align} }
\def\bf{
\begin{figure} }
  \def\ef{
\end{figure} }
\def\bea{
\begin{eqnarray} }
  \def\eea{
\end{eqnarray} }
\def\bse{
\begin{subaligns} }
  \def\ese{
\end{subaligns} }

\def\P01{ P_{\text{0}\rightarrow \text{1}} }

\def\be{ \begin{equation}}
\def\ee{ \end{equation}}
\def\bse{  \begin{subequations}}
\def\ese{  \end{subequations}}
\def\bea#1\ea{\begin{align}#1\end{align}}

\def\bi{\begin{itemize}}
\def\ei{\end{itemize}}

\def\bt{\begin{tabular}}
\def\et{\end{tabular}}

\def\rms{\Lambda}

\usepackage{hyperref}
\hypersetup{hidelinks}
\usepackage[final]{microtype}
\ifPDFTeX
  \microtypesetup{protrusion=true,expansion=true,tracking=true,kerning=true,spacing=true}
\fi
\makeatletter
\long\def\@makecaption#1#2{%
  \vskip\abovecaptionskip
  \par\noindent
  \begin{minipage}{\linewidth}
    \normalsize\textbf{#1}.\ \justifying #2
  \end{minipage}\par
  \vskip\belowcaptionskip
}
\makeatother

\begin{document}

\author{Branislav Ilikj}
\address{Center for Quantum Technologies, Department of Physics, Sofia University, James Bourchier 5 blvd, 1164 Sofia, Bulgaria}
\author{Thomas Zanon-Willette}
\address{Sorbonne Université, CNRS, MONARIS, UMR 8233, F-75005 Paris, France}
\author{Nikolay V. Vitanov}
\address{Center for Quantum Technologies, Department of Physics, Sofia University, James Bourchier 5 blvd, 1164 Sofia, Bulgaria}

\title{Ramsey Interferometry in Wigner--Majorana Qudits}
\date{\today }

\begin{abstract}
Ramsey interferometry estimates a detuning from the phase accumulated between two separated interaction zones, with a characteristic resolution set by the interrogation time $\tau$. 
We propose and analyze a single-qudit extension of Ramsey interferometry based on Wigner--Majorana (WM) spin-$j$ dynamics, where the internal levels of one quantum system form a multipath interferometer. 
The enhancement is not a generic consequence of replacing qubit $\pi/2$ pulses by abstract qudit gates: ideal $\mathrm{QFT}_D$ and $\sqrt{X_D}$ Ramsey sequences do not produce a clean densification of the central Ramsey fringe under the population readouts considered here. 
Instead, in physical manifolds that realize the WM coupling structure, the effect arises from coherences between separated ladder states that are created and recombined by a single near-resonant drive in each Ramsey zone.

The clearest result is obtained for a qutrit. 
Preparing the central state of a spin-1 WM manifold and measuring its return probability gives the ideal-pulse Ramsey signal
$P_3(\Delta)=\cos^2(\Delta\tau)$, to be compared with the qubit signal $P_2(\Delta)=\cos^2(\Delta\tau/2)$. 
Thus the qutrit central fringe is compressed by a factor of two at fixed $\tau$, and the maximal central-fringe slope is doubled in the ideal limit while the contrast remains ideally unity. 
For higher-dimensional WM manifolds, both odd and even $D$, the central response becomes sharper with increasing dimension, but the useful population signal is distributed over several output channels. 
We therefore introduce a fixed scalar readout based on nearest-neighbor shoulder populations and quantify the resulting resolution-contrast trade-off through the central-fringe slope $S_D$ and contrast $C_D$.

We also examine robustness against diagonal phase noise motivated by probe-shift fluctuations. 
Under this projector-type common-mode dephasing model, the selected WM readouts are less contrast-sensitive than the corresponding qubit readout for the parameter set considered.
This robustness is symmetry-selective rather than universal: for linear Zeeman dephasing ($L=J_z$), the same large $m$-separation that compresses the fringe also enhances dephasing, so higher-dimensional WM readouts become increasingly sensitive to it rather than protected.
These results identify the WM qutrit as a particularly practical operating point for enhanced Ramsey spectroscopy at fixed interrogation time, while higher-dimensional WM qudits offer additional slope enhancement at the cost of reduced contrast.
\end{abstract}

\maketitle

\section{Introduction}

Ramsey interferometry \cite{Ramsey1950,Ramsey1956} is a cornerstone of high-resolution spectroscopy and precision metrology \cite{Barbieri2022}.
In its standard two-level form, a system prepared in a basis state is driven by two separated
$\pi/2$ pulses, producing interference fringes as a function of detuning $\Delta$ whose characteristic
feature width scales as $1/\tau$, where $\tau$ is the free-evolution (interrogation) time.
This simple scaling underlies a wide range of applications, from frequency standards and atomic
clocks~\cite{Ludlow2015,DereviankoKatori2011} to sensors and coherent spectroscopy, and it has motivated a large body of work on improving
accuracy and robustness by tailoring the two-pulse sequence like, hyper-Ramsey, autobalanced, Ramsey-comb, and geometric Ramsey variants, several aimed at suppressing systematic probe shifts \cite{Morgenweg2014,Yudin2018GABRS,Madasu2024,ZanonWillette2015,Hobson2016}.

A complementary question --- central to this work --- is whether one can increase the \emph{detuning response}
at fixed interrogation time \emph{without} invoking multipartite entanglement, squeezing, or longer
coherence times. Such resource-based routes to precision, together with their fundamental limits, have been analyzed extensively~\cite{gefen2019overcoming,abiuso2025fundamental,Zhou2023,olivares2025quantum}.
One natural resource is the internal multilevel structure of many experimental
platforms: hyperfine and Zeeman manifolds in atoms and ions~\cite{Leibfried2003,Saffman2010RMP,Lindon2023}, rotational manifolds in molecules,
and engineered multilevel circuits. 
In the language of quantum information, such systems are single \emph{qudits} with Hilbert-space dimension $D>2$ \cite{Wang2021,Godfrin2018}. While qudits are widely discussed as a route to larger computational spaces, they are often assumed to require substantially increased control complexity. 
In contrast, spectroscopy routinely accesses multilevel manifolds using a small number of near-resonant fields. 
The central aim of this paper is to identify when and how a \emph{single} driven qudit can act as a practical multipath Ramsey interferometer that yields a steeper central response to detuning than the qubit baseline, while preserving a control workload comparable to standard two-level Ramsey spectroscopy.

We focus on a physically ubiquitous and analytically transparent class of qudits: manifolds whose
driven dynamics obey an embedded SU(2) symmetry (Wigner--Majorana or spin-$j$ dynamics) \cite{Wigner1959},
with $D=2j+1$ and nearest-neighbor ladder couplings. 
This WM structure arises naturally in many atomic and ionic settings and can also be realized effectively in $\Lambda$-type Raman implementations.
Within this framework, a Ramsey sequence built from two identical driven segments separated by
free evolution implements a multipath interference process over the WM ladder states. 
In physical manifolds that naturally realize the WM ladder couplings, this Ramsey interrogation can be implemented with a \emph{single} near-resonant driving field in each interaction zone, just as in the qubit case. 
The multilevel structure is therefore exploited as an interferometric resource rather than as an added control burden.

A key point is that not every ``qudit-generalized'' Ramsey idea leads to genuine fringe densification
in a physically meaningful readout. 
As an instructive baseline, we analyze interrogations in which the two $\pi/2$ pulses are replaced by idealized qudit gates such as $\mathrm{QFT}_D$ or $\sqrt{X_D}$ separated by free evolution. 
In the ideal-gate model --- where these operations are treated as $\Delta$-independent unitaries
and detuning enters only through the free-evolution segment --- these gate-based protocols do not increase the
density of the central Ramsey features in the resulting population signal. 
By contrast, in WM manifolds the \emph{driven} multilevel dynamics itself generates multipath phases that can sharpen the central response.

To compare different dimensions on equal footing, we adopt an operational viewpoint: the experiment produces
projective outcomes in a fixed basis, and one constructs from them a single scalar readout $P_D(\Delta)$ that is
used to infer $\Delta$. For a qutrit ($D=3$) a natural choice is the return probability to the central WM state,
whereas for $D>3$ the dominant near-central oscillatory response is typically carried by two symmetric
nearest-neighbor ``shoulder'' channels, motivating a binned shoulder-sum readout.
We quantify performance of the central Ramsey feature by two directly extracted metrics:
(i) the maximal absolute slope $S_D$ of the central fringe, and (ii) the corresponding contrast $C_D$.
These two numbers encapsulate a practical resolution-contrast trade-off under a fixed measurement definition:
a steeper slope improves detuning sensitivity, while reduced contrast degrades it through projection noise.
(Where useful, one may connect these quantities to standard estimation theory by viewing the binned readout
as a two-outcome probability, for which the classical Fisher information scales as
$F_{\mathrm{cl}}(\Delta)\propto (\partial_\Delta P_D)^2/[P_D(1-P_D)]$.)

Within this unified readout framework, we find that WM qudits can exhibit a marked enhancement of the central
detuning response at fixed $\tau$. The most transparent result is obtained for the qutrit:
an analytic expression for the WM Ramsey signal shows that, in the ideal-pulse limit, the central return probability is
$P_3(\Delta)=\cos^2(\Delta\tau)$, i.e., the fringe density is doubled compared with the qubit baseline
$P_2(\Delta)=\cos^2(\Delta\tau/2)$. 
For higher dimensions ($D>3$), numerical simulations of WM Ramsey sequences
show an approximately linearly increasing central slope with $D$ under suitable preparation and readout choices, while
the contrast generally decreases, yielding a dimension-dependent optimum. 
In the parameter regime studied here, the qutrit emerges as a particularly favorable ``sweet spot'': it provides roughly a twofold slope enhancement with essentially no loss of contrast, whereas larger $D$ yields further fringe compression at the cost of reduced visibility.

Finally, since any practical advantage must persist under realistic imperfections, we assess robustness to an
experimentally motivated class of diagonal phase noise associated with probe-shift (AC Stark shift) fluctuations.
Using a GKSL description with a diagonal jump operator that induces arm-selective phase diffusion in the WM basis,
we find that, under this specific common-mode diagonal-noise model, the qutrit WM interference pattern is less
contrast-sensitive than the corresponding qubit readout. 
We emphasize that this robustness is \emph{not} universal for arbitrary diagonal dephasing, but is instead tied to the symmetry structure of the specific probe-shift channel considered. The linear Zeeman channel ($L=J_z$) behaves oppositely: because it distinguishes states by their magnetic quantum number, the same wide $m$-separation that sharpens the fringe also accelerates its dephasing, so the higher-dimensional readouts grow progressively more fragile as $D$ increases.

Beyond the single-qudit route pursued here, a range of complementary strategies enhance frequency sensitivity: squeezing and entanglement-based quantum metrology~\cite{Wineland1992,Giovannetti2011,Demkowicz2012}, adaptive Bayesian Ramsey measurements~\cite{McMichael2021}, and other high-resolution spectroscopy techniques~\cite{Mihov2024}, all building on the foundational magnetic-resonance method of Rabi \emph{et al.}~\cite{Rabi1937}. Most closely related to the present work, three-level Ramsey interferometry has recently been proposed as a resource for enhanced metrology~\cite{ZhouLi2026}. Related multipath and non-Hermitian sensing platforms have likewise been explored~\cite{Wiersig2020,Xiao2024}.

The paper is organized as follows.
In Sec.~\ref{sec:qubit-ramsey} we review qubit Ramsey interferometry and define the readout and central-feature metrics used throughout. 
In Sec.~\ref{sec:qudit_ramsey} we introduce gate-based qudit Ramsey baselines based on $\mathrm{QFT}_D$ and $\sqrt{X_D}$. 
In Sec.~\ref{sec:wm} we analyze WM Ramsey interrogations, including an analytic treatment of the qutrit and a systematic comparison of odd and even dimensions using the appropriate readouts. 
In Sec.~\ref{sec:decoherence} we study decoherence effects in the WM setting, focusing on diagonal phase noise motivated by probe-shift fluctuations. 
Sec.~\ref{sec:discussion} discusses the resulting resolution--contrast trade-offs and the practical operating regimes suggested by our simulations, and Sec.~\ref{sec:conclusion} concludes the paper.


\section{Standard qubit Ramsey interferometry}
\label{sec:qubit-ramsey}

We first recall standard Ramsey interferometry for a two-level system, which will serve as the reference point for the qudit protocols discussed below. The objective of a Ramsey experiment is to estimate an unknown detuning $\Delta$ between the applied driving field and the qubit transition frequency. The detuning is inferred from the population oscillations generated by two separated $\pi/2$ pulses and a free-evolution interval of duration $\tau$.

We work in the rotating frame and set $\hbar=1$. The qubit state is written as $|\psi(t)\rangle=c_1(t)|1\rangle+c_2(t)|2\rangle$, and its dynamics obey $\mathrm{i}\,d|\psi(t)\rangle/dt=H_2(t)|\psi(t)\rangle$. Within the rotating-wave approximation, a rectangular pulse with Rabi frequency $\Omega$~\cite{Rabi1937} and detuning $\Delta$ is described by the Hamiltonian
\begin{equation}
H_2
=
\frac{1}{2}
\begin{pmatrix}
-\Delta & \Omega \\
\Omega & \Delta
\end{pmatrix}
=
\frac{1}{2}
\left(
\Omega\sigma_x-\Delta\sigma_z
\right).
\end{equation}
The corresponding pulse propagator is $R_2(T,\Delta)=\exp(-\mathrm{i}H_2T)$. Introducing the generalized Rabi frequency $\rms=\sqrt{\Omega^2+\Delta^2}$, the finite-duration pulse propagator can be written as
\begin{equation}
R_2(T,\Delta)
=
\begin{pmatrix}
c+\mathrm{i}\frac{\Delta}{\rms}s
&
-\mathrm{i}\frac{\Omega}{\rms}s
\\
-\mathrm{i}\frac{\Omega}{\rms}s
&
c-\mathrm{i}\frac{\Delta}{\rms}s
\end{pmatrix},
\end{equation}
where $c=\cos(\rms T/2)$ and $s=\sin(\rms T/2)$. The pulse duration is calibrated on resonance as a $\pi/2$ pulse, $\Omega T=\pi/2$. However, for nonzero detuning the actual pulse evolution is governed by $\rms$, not by $\Omega$. Thus the detuning affects both the phase accumulated during each pulse and the effective pulse area.

During the dark time, the Rabi coupling is absent and the free-evolution propagator is
\begin{equation}\label{F2}
F_2(\tau,\Delta)
=
\begin{pmatrix}
e^{\mathrm{i}\phi} & 0 \\
0 & e^{-\mathrm{i}\phi}
\end{pmatrix},
\end{equation}
where $\phi=\Delta\tau/2$ is the dark-time phase.

There are two natural ways to choose the second Ramsey pulse. The standard choice is to repeat the first pulse. The corresponding finite-pulse Ramsey propagator is
\[
U_2^{(\mathrm{rep})}(\Delta)
=
R_2(T,\Delta)F_2(\tau,\Delta)R_2(T,\Delta).
\]
For an initial state $|1\rangle$, the probability to find the qubit in $|2\rangle$ after the sequence is
\[
P_2^{(\mathrm{rep})}(\Delta)
=
\left|
\langle 2|U_2^{(\mathrm{rep})}(\Delta)|1\rangle
\right|^2 .
\]
Using the propagators above, one obtains
\begin{equation}
P_2^{(\mathrm{rep})}(\Delta)
=
4\frac{\Omega^2}{\rms^2}s^2
\left[
c\cos\phi
-
\frac{\Delta}{\rms}s\sin\phi
\right]^2 .
\label{eq:finite-pulse-qubit-ramsey-repeated}
\end{equation}
This expression includes the effect of the same unknown detuning during both rectangular Ramsey pulses and during the free-evolution interval.

In the ideal-pulse limit, the detuning-induced pulse distortion is neglected. This requires $|\Delta|/\Omega\ll 1$, together with a pulse duration short compared with the interrogation time. The resonant $\pi/2$ pulse then gives $c=s=1/\sqrt{2}$. Equation~\eqref{eq:finite-pulse-qubit-ramsey-repeated} reduces to
\begin{equation}
P_2^{(\mathrm{rep})}(\Delta)
=
\cos^2\left(\frac{\Delta\tau}{2}\right).
\label{eq:ideal-qubit-ramsey-repeated}
\end{equation}
Thus the repeated-pulse Ramsey sequence gives a bright central fringe~\cite{Ramsey1950,Ramsey1956}.

A second useful choice is to make the second Ramsey pulse the inverse of the first one. The corresponding propagator is
\[
U_2^{(\mathrm{inv})}(\Delta)
=
R_2^\dagger(T,\Delta)F_2(\tau,\Delta)R_2(T,\Delta).
\]
This choice cancels the first pulse exactly when $\tau=0$, because the sequence then reduces to the identity. The transition probability is
\[
P_2^{(\mathrm{inv})}(\Delta)
=
\left|
\langle 2|U_2^{(\mathrm{inv})}(\Delta)|1\rangle
\right|^2 .
\]
For finite rectangular pulses, one obtains
\begin{equation}
P_2^{(\mathrm{inv})}(\Delta)
=
4\frac{\Omega^2}{\rms^2}s^2
\left[
c^2
+
\frac{\Delta^2}{\rms^2}s^2
\right]
\sin^2\phi .
\label{eq:finite-pulse-qubit-ramsey-inverse}
\end{equation}
In the ideal-pulse limit, Eq.~\eqref{eq:finite-pulse-qubit-ramsey-inverse} reduces to
\begin{equation}
P_2^{(\mathrm{inv})}(\Delta)
=
\sin^2\left(\frac{\Delta\tau}{2}\right).
\label{eq:ideal-qubit-ramsey-inverse}
\end{equation}
Thus the inverse-pulse Ramsey sequence gives the complementary dark central fringe.

The inverse pulse can be generated without knowing the detuning. Let $Y$ denote the Pauli $y$ operator. Since $Y\sigma_xY=-\sigma_x$ and $Y\sigma_zY=-\sigma_z$, one finds $YH_2Y=-H_2$. Therefore,
\[
YR_2(T,\Delta)Y
=
R_2^\dagger(T,\Delta).
\]
The inverse-pulse Ramsey sequence can then be implemented as
\[
U_2^{(\mathrm{inv})}(\Delta)
=
YR_2(T,\Delta)YF_2(\tau,\Delta)R_2(T,\Delta).
\]
If $Y$ is implemented as a physical $\pi$ rotation about the $y$ axis, this equality holds up to an irrelevant global phase. The important point is that the $Y$ sandwich inverts both the drive term and the unknown detuning term. Hence the construction does not require prior knowledge of $\Delta$.

The repeated-pulse and inverse-pulse Ramsey signals differ by their readout phase. In the ideal limit, the first gives a bright central fringe,
\begin{equation}\label{P2rep}
P_2^{(\mathrm{rep})}(\Delta)
=
\cos^2\left(\frac{\Delta\tau}{2}\right),
\end{equation}
whereas the second gives a dark central fringe,
\begin{equation}\label{P2inv}
P_2^{(\mathrm{inv})}(\Delta)
=
\sin^2\left(\frac{\Delta\tau}{2}\right).
\end{equation}
Both ideal signals have the same contrast, fringe width, maximal slope, and Fisher information. Thus the inverse-pulse sequence changes the Ramsey readout convention but does not improve the fundamental Ramsey resolution. For finite pulses, however, Eqs.~\eqref{eq:finite-pulse-qubit-ramsey-repeated} and \eqref{eq:finite-pulse-qubit-ramsey-inverse} are not identical up to a phase shift: the repeated-pulse sequence contains finite-pulse phase and area corrections, while the inverse-pulse sequence cancels the pulse evolution exactly at zero dark time.

To compare qubit and qudit Ramsey protocols on the same footing, we use a scalar population readout $P_D(\Delta)$ and characterize its central Ramsey feature by the central-fringe slope $S_D$, the contrast $C_D$, and the width $W_D$. The precise definitions, together with the corresponding Fisher-information sensitivity, are given in Appendix~\ref{app:ramsey-metrics}. For either ideal qubit Ramsey convention, these quantities are $C_2=1$, $S_2=\tau/2$, and $W_2=\pi/\tau$. For the finite-pulse signals in Eqs.~\eqref{eq:finite-pulse-qubit-ramsey-repeated} and \eqref{eq:finite-pulse-qubit-ramsey-inverse}, the corresponding values should instead be extracted directly from the full detuning-dependent expressions.

\section{Gate-based qudit Ramsey baselines}\label{sec:qudit_ramsey}\label{sec:qft}

\begin{figure}[t]
  \centerline{\includegraphics[width=\columnwidth]{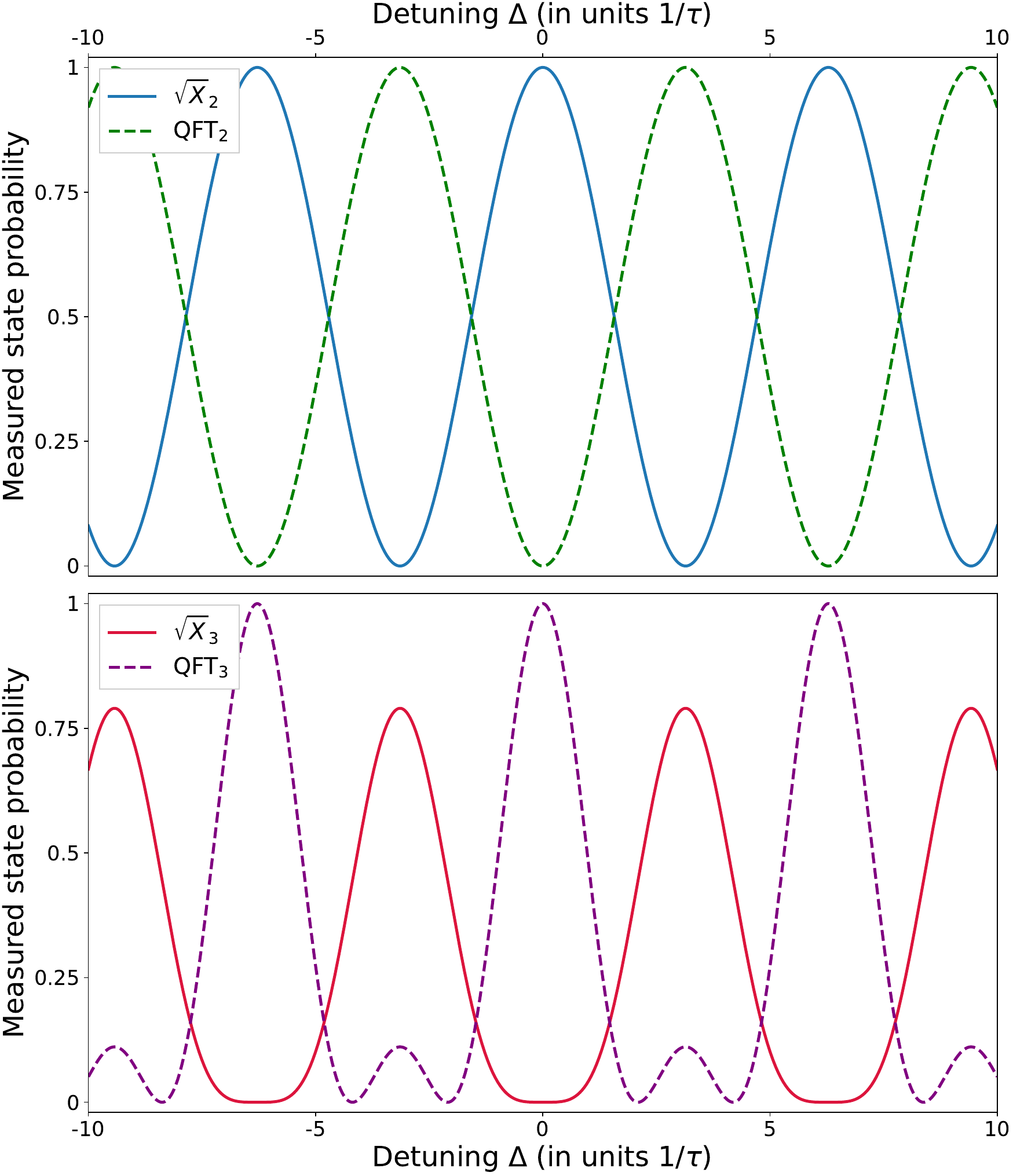}}
  \caption{Ideal-gate Ramsey baselines for qubit and qutrit readouts. Top: qubit oscillations for $\sqrt{X}_2$ (solid) and $\mathrm{QFT}_2$ (dashed). Bottom: qutrit central-state oscillations for $\sqrt{X}_3$ (solid) and $\mathrm{QFT}_3$ (dashed). Parameters: $\tau=1$, $\Omega T=\pi/2$, and $T=0.1$.}
  \label{fig:qft}
\end{figure}

Before introducing Wigner--Majorana (WM) Ramsey interferometry, we first examine the most direct qudit-gate analogue of the qubit Ramsey sequence. The purpose of this section is not to propose an optimal qudit protocol, but to test a natural idea: replace the qubit Ramsey pulse, or equivalently the qubit Hadamard operation, by its standard qudit counterpart.

This provides an important baseline. If the enhancement discussed later were simply a consequence of using a larger Hilbert space, then a standard qudit gate such as \(\mathrm{QFT}_D\) should already produce a clean compression of the Ramsey fringe. We show below that this is not the case. The ideal-gate qudit construction creates multilevel interference, but it does not isolate the same useful outer-state coherence that appears naturally in the WM qutrit.

We consider an equally spaced \(D\)-level ladder with basis states \(|1\rangle,\ldots,|D\rangle\). During the dark time, the detuning-dependent free evolution is
\begin{equation}
F_D(\tau,\Delta)
=
\sum_{n=1}^{D}
\exp\left[
-\mathrm{i}\left(n-\frac{D+1}{2}\right)\Delta\tau
\right]
|n\rangle\langle n|.
\label{eq:qudit_dark_evolution}
\end{equation}
For compactness, we write \(\theta=\Delta\tau\). In this ideal-gate model, the Ramsey pulses are treated as instantaneous, detuning-independent unitary gates. Thus the unknown detuning is accumulated only during the dark interval. This is different from the finite-pulse qubit treatment in Sec.~\ref{sec:qubit-ramsey}, where the same detuning acts during the driven pulses as well.

Let \(G_D\) be the first qudit Ramsey gate. In analogy with Sec.~\ref{sec:qubit-ramsey}, there are two natural choices for the second pulse. The first is to repeat the same gate,
\begin{equation}
U_D^{(\mathrm{rep})}(\Delta)
=
G_D F_D(\tau,\Delta)G_D .
\label{eq:qudit_rep_sequence}
\end{equation}
The second is to apply the inverse gate,
\begin{equation}
U_D^{(\mathrm{inv})}(\Delta)
=
G_D^\dagger F_D(\tau,\Delta)G_D .
\label{eq:qudit_inv_sequence}
\end{equation}
The inverse convention has a useful reference property: for zero dark time it reduces to the identity. The repeated convention instead reduces to \(G_D^2\). This is the same distinction as in the qubit case of Sec.~\ref{sec:qubit-ramsey}: the inverse sequence is a clean prepare--accumulate--unprepare interferometer, while the repeated sequence corresponds more directly to applying two identical Ramsey zones. As in the qubit case, however, this distinction by itself is not a resolution advantage. It changes the readout convention and the zero-phase reference.

For an initial state \(|p\rangle\) and a measured state \(|q\rangle\), the repeated-gate Ramsey amplitude is
\begin{equation}
A_{p\rightarrow q}^{(\mathrm{rep})}(\Delta)
=
\sum_{n=1}^{D}
(G_D)_{qn}
(G_D)_{np}
\exp\left[
-\mathrm{i}\left(n-\frac{D+1}{2}\right)\theta
\right].
\label{eq:qudit_rep_amplitude}
\end{equation}
The inverse-gate Ramsey amplitude is
\begin{equation}
A_{p\rightarrow q}^{(\mathrm{inv})}(\Delta)
=
\sum_{n=1}^{D}
(G_D)_{nq}^{*}
(G_D)_{np}
\exp\left[
-\mathrm{i}\left(n-\frac{D+1}{2}\right)\theta
\right].
\label{eq:qudit_inv_amplitude}
\end{equation}
The measured population signals are obtained from \(P_{p\rightarrow q}^{(\mathrm{rep})}(\Delta)=|A_{p\rightarrow q}^{(\mathrm{rep})}(\Delta)|^2\) and \(P_{p\rightarrow q}^{(\mathrm{inv})}(\Delta)=|A_{p\rightarrow q}^{(\mathrm{inv})}(\Delta)|^2\).

Equations \eqref{eq:qudit_rep_amplitude} and \eqref{eq:qudit_inv_amplitude} show the basic issue. A qudit gate can distribute amplitude over many levels, but the Ramsey signal depends on how the resulting detuning phases are recombined into the chosen output channel. A multilevel superposition alone does not guarantee a useful compressed fringe.

\subsection{The qudit Fourier transform as the Hadamard replacement}

The most standard qudit generalization of the Hadamard gate is the qudit quantum Fourier transform~\cite{Shor1997,Coppersmith1994,Pudda2024}. We define it by
\begin{equation}
\mathrm{QFT}_D |n\rangle
=
\frac{1}{\sqrt{D}}
\sum_{k=1}^{D}
\omega^{(n-1)(k-1)}
|k\rangle ,
\label{eq:qft_def_secIII}
\end{equation}
where \(\omega=e^{2\pi\mathrm{i}/D}\). For \(D=2\), \(\mathrm{QFT}_2\) is the usual qubit Hadamard gate~\cite{Hadamard1893,Sylvester1867,Yurtalan2020WH}, up to basis phases. It is therefore the natural first gate-based replacement to try in a qudit Ramsey sequence.

We now set \(G_D=\mathrm{QFT}_D\). Substituting Eq. \eqref{eq:qft_def_secIII} into the inverse-gate amplitude gives
\begin{align}
A_{p\rightarrow q,QFT}^{(\mathrm{inv})}(\Delta)
&=
\frac{1}{D}
\sum_{r=0}^{D-1}
\exp\left[
-\mathrm{i}\frac{2r+1-D}{2}\theta
\right] \notag\\
&\times\exp\left[
\frac{2\pi\mathrm{i} r(p-q)}{D}
\right].
\label{eq:qft_inv_amplitude}
\end{align}
This is a finite Fourier sum. The corresponding signal is
\begin{equation}
P_{p\rightarrow q,QFT}^{(\mathrm{inv})}(\Delta)
=
\frac{1}{D^2}
\frac{
\sin^2\left[
\frac{D}{2}
\left(
\theta-\frac{2\pi(p-q)}{D}
\right)
\right]
}{
\sin^2\left[
\frac{1}{2}
\left(
\theta-\frac{2\pi(p-q)}{D}
\right)
\right]
}.
\label{eq:qft_inv_signal}
\end{equation}
The value at a removable singularity is understood by continuity. For the repeated-gate sequence, the corresponding amplitude is
\begin{align}
A_{p\rightarrow q,QFT}^{(\mathrm{rep})}(\Delta)
&=
\frac{1}{D}
\sum_{r=0}^{D-1}
\exp\left[
-\mathrm{i}\frac{2r+1-D}{2}\theta
\right] \notag\\
&\times \exp\left[
\frac{2\pi\mathrm{i} r(q+p-2)}{D}
\right].
\label{eq:qft_rep_amplitude}
\end{align}
The repeated-gate signal is
\begin{equation}
P_{p\rightarrow q,QFT}^{(\mathrm{rep})}(\Delta)
=
\frac{1}{D^2}
\frac{
\sin^2\left[
\frac{D}{2}
\left(
\theta-\frac{2\pi(q+p-2)}{D}
\right)
\right]
}{
\sin^2\left[
\frac{1}{2}
\left(
\theta-\frac{2\pi(q+p-2)}{D}
\right)
\right]
}.
\label{eq:qft_rep_signal}
\end{equation}
These expressions are the ideal-gate qudit analogues of the two qubit Ramsey conventions in Sec.~\ref{sec:qubit-ramsey}.

For \(D=2\), \(QFT_2\) is self-inverse. Therefore the repeated and inverse conventions coincide at the level of the ideal Hadamard sequence. With return readout, the signal is \(\cos^2(\theta/2)\); with the other output channel, it is \(\sin^2(\theta/2)\). These are the usual complementary qubit Ramsey fringes. Thus the \(D=2\) limit reproduces ordinary Ramsey behavior, up to the choice of readout phase.

\subsection{Qutrit QFT Ramsey signal}

We now apply this construction to the qutrit, \(D=3\). The central state is \(|2\rangle\). For the inverse-gate convention, the natural scalar readout is the return probability to the same state. Setting \(p=q=2\) in Eq. \eqref{eq:qft_inv_signal} gives
\begin{equation}
P_{2\rightarrow 2,QFT_3}^{(\mathrm{inv})}(\Delta)
=
\frac{1}{9}
\left[
1+2\cos(\theta)
\right]^2 .
\label{eq:qft3_inv_signal}
\end{equation}
Equivalently,
\begin{equation}
P_{2\rightarrow 2,QFT_3}^{(\mathrm{inv})}(\Delta)
=
\frac{1}{9}
\left[
4\cos^2\left(\frac{\Delta\tau}{2}\right)-1
\right]^2 .
\label{eq:qft3_inv_signal_alt}
\end{equation}
This signal has a central maximum at \(\Delta=0\), but it is not the clean WM qutrit signal \(P_3(\Delta)=\cos^2(\Delta\tau)\). Expanding Eq. \eqref{eq:qft3_inv_signal}, one sees that it contains several Fourier components rather than a single doubled Ramsey phase. The QFT gate has created a multilevel interference pattern, but it has not isolated the outer-state coherence in the simple two-path form needed for a clean factor-of-two fringe compression.

For the repeated-gate convention, the zero-dark-time sequence is \(QFT_3^2\), not the identity. With the convention of Eq. \eqref{eq:qft_def_secIII}, \(QFT_3^2|2\rangle=|3\rangle\). Thus the direct analogue of the bright repeated-pulse readout is the transition probability from the central input state \(|2\rangle\) to \(|3\rangle\). Setting \(p=2\) and \(q=3\) in Eq. \eqref{eq:qft_rep_signal} gives
\begin{equation}
P_{2\rightarrow 3,QFT_3}^{(\mathrm{rep})}(\Delta)
=
\frac{1}{9}
\left[
1+2\cos(\theta)
\right]^2 .
\label{eq:qft3_rep_bright_signal}
\end{equation}
Thus, with the output channel chosen to match the zero-dark-time action of \(QFT_3^2\), the repeated-gate convention gives the same central trace as the inverse-gate return signal. This is analogous to the qubit situation in the limited sense that the repeated and inverse conventions differ mainly by the readout reference.

However, if one insists on measuring the central state after the repeated \(QFT_3\) sequence, the signal is instead
\begin{equation}
P_{2\rightarrow 2,QFT_3}^{(\mathrm{rep})}(\Delta)
=
\frac{
\sin^2\left(\frac{3\theta}{2}\right)
}{
9\sin^2\left(\frac{\theta}{2}-\frac{2\pi}{3}\right)
}.
\label{eq:qft3_rep_central_signal}
\end{equation}
This illustrates the main weakness of the ideal-gate approach. The output signal is strongly tied to the chosen gate phases and the chosen measurement channel. The qudit Fourier transform is the standard replacement of the Hadamard gate, but the resulting qutrit Ramsey pattern is not a simple, high-contrast signal of the form \(\cos^2(\Delta\tau)\).

The reason is physical. The QFT gate creates a Fourier superposition over the ladder, but it does not specifically prepare the two outer qutrit states as the two arms of the interferometer. In contrast, the WM qutrit pulse discussed in the next section maps the central state directly onto the two outer states. These two arms have opposite detuning shifts and therefore acquire a relative phase \(2\Delta\tau\). The second WM pulse recombines precisely this outer-state coherence back onto the central state, producing the ideal signal \(P_3(\Delta)=\cos^2(\Delta\tau)\).

The QFT construction is therefore a useful negative result. It shows that the enhancement is not a generic consequence of replacing a qubit by a qutrit, nor of replacing a Hadamard by a qudit Fourier transform. The useful enhancement requires the physical WM coupling structure.

\subsection{Comment on the square-root shift gate}

One may also consider the qudit square-root shift gate \(\sqrt{X_D}\), defined by \(\sqrt{X_D}=\mathrm{QFT}_D\sqrt{Z_D}\mathrm{QFT}_D^\dagger\). This is another natural ideal-gate analogue of a qubit \(\pi/2\) operation, especially if one thinks of Ramsey interferometry in terms of resonant \(\pi/2\) rotations rather than Hadamard gates.

For the purposes of the present argument, however, \(\sqrt{X_D}\) does not need to be treated in the main text. Substituting \(G_D=\sqrt{X_D}\) into Eqs. \eqref{eq:qudit_rep_amplitude} and \eqref{eq:qudit_inv_amplitude} leads to the same qualitative conclusion: the resulting qutrit population traces are channel-dependent multilevel interference signals, not the clean WM response \(P_3(\Delta)=\cos^2(\Delta\tau)\). We therefore use \(\mathrm{QFT}_D\) as the representative ideal-gate baseline and reserve other gate choices, such as \(\sqrt{X_D}\), for optional checks.

\section{Wigner--Majorana Ramsey interferometry}
\label{sec:wm}

\begin{figure}[!t]
  \centerline{\includegraphics[width=\linewidth]{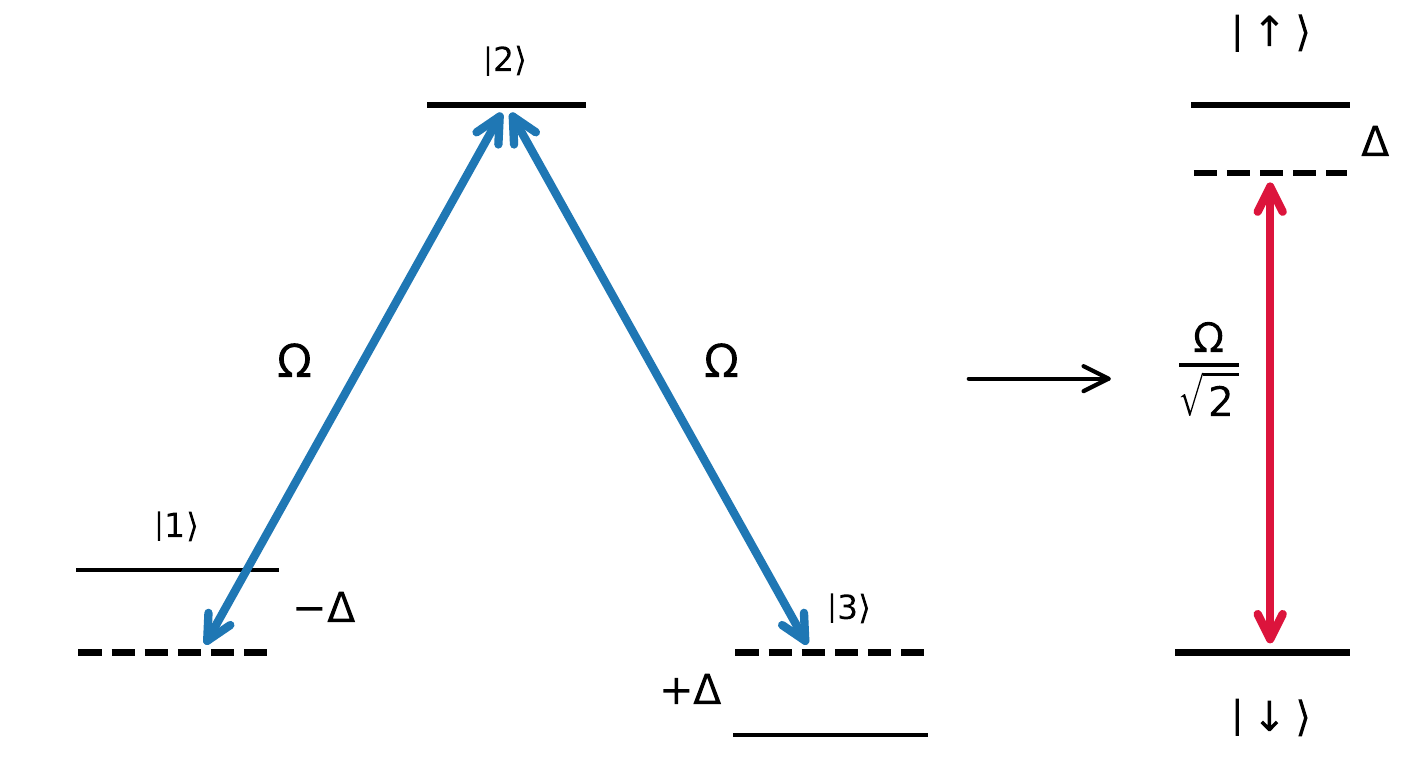}}
  \caption{Wigner--Majorana (WM) decomposition of a $\Lambda$-type qutrit system. Left: three-state configuration ($m=-1,0,1$). Right: effective two-state system ($m=-\frac{1}{2}, \frac{1}{2}$) obtained after WM decomposition.}
  \label{fig:states}
\end{figure}
\begin{figure}[!t]
  \centerline{\includegraphics[width=\linewidth]{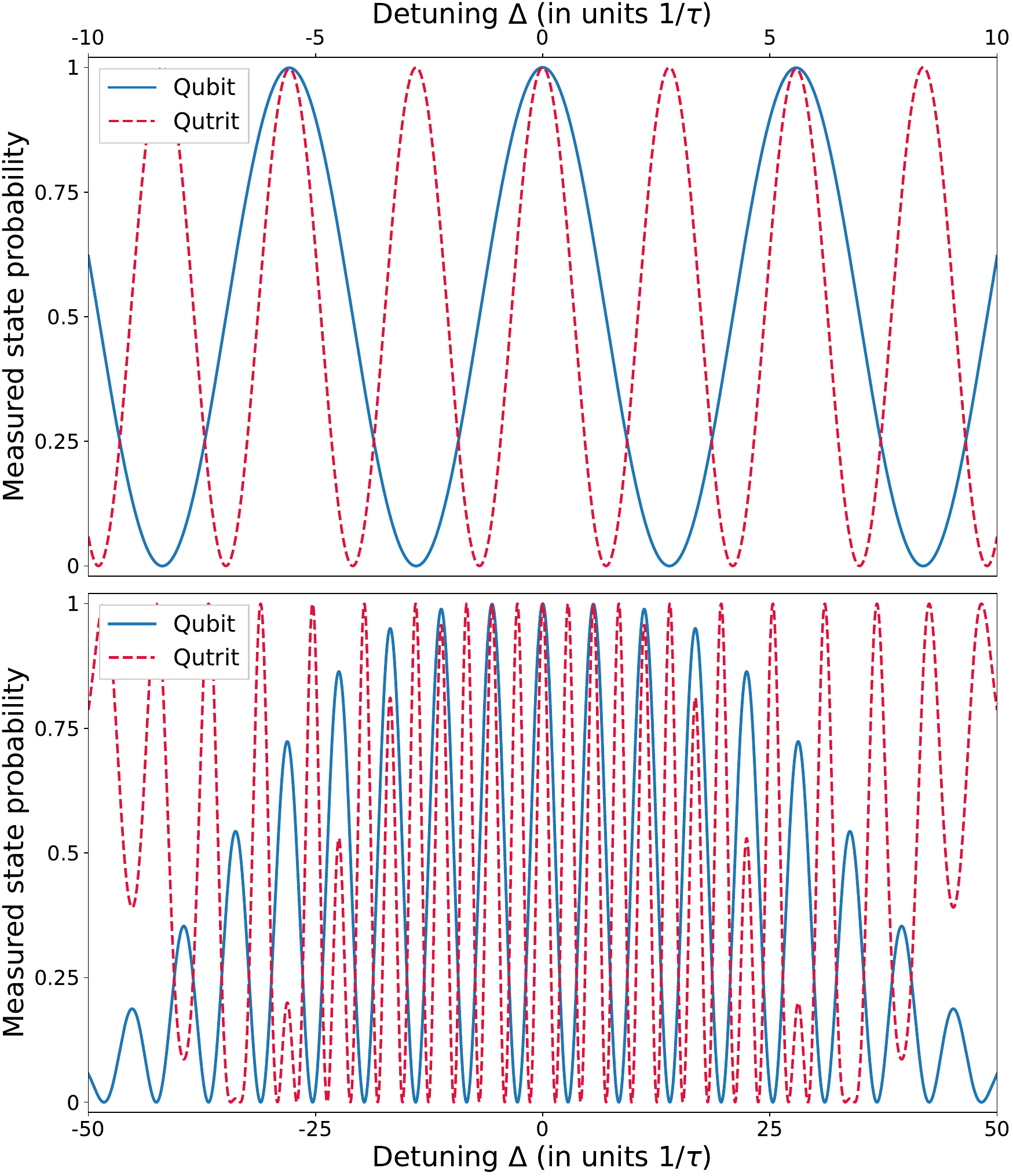}}
  \caption{Qubit and qutrit WM Ramsey oscillations. The solid curve shows the qubit reference $P_2(\Delta)=\cos^2(\Delta\tau/2)$, while the dashed curve shows the ideal qutrit WM response $P_3(\Delta)=\cos^2(\Delta\tau)$. Top panel: $\Delta\in[-10,10]$ units of $1/\tau$; bottom panel: $\Delta\in[-50,50]$ units of $1/\tau$. Other parameters are identical to those used in Fig.~\ref{fig:qft}.}
  \label{fig:qutrit_oscillations}
\end{figure}
\begin{figure}[!t]
  \centerline{\includegraphics[width=\linewidth]{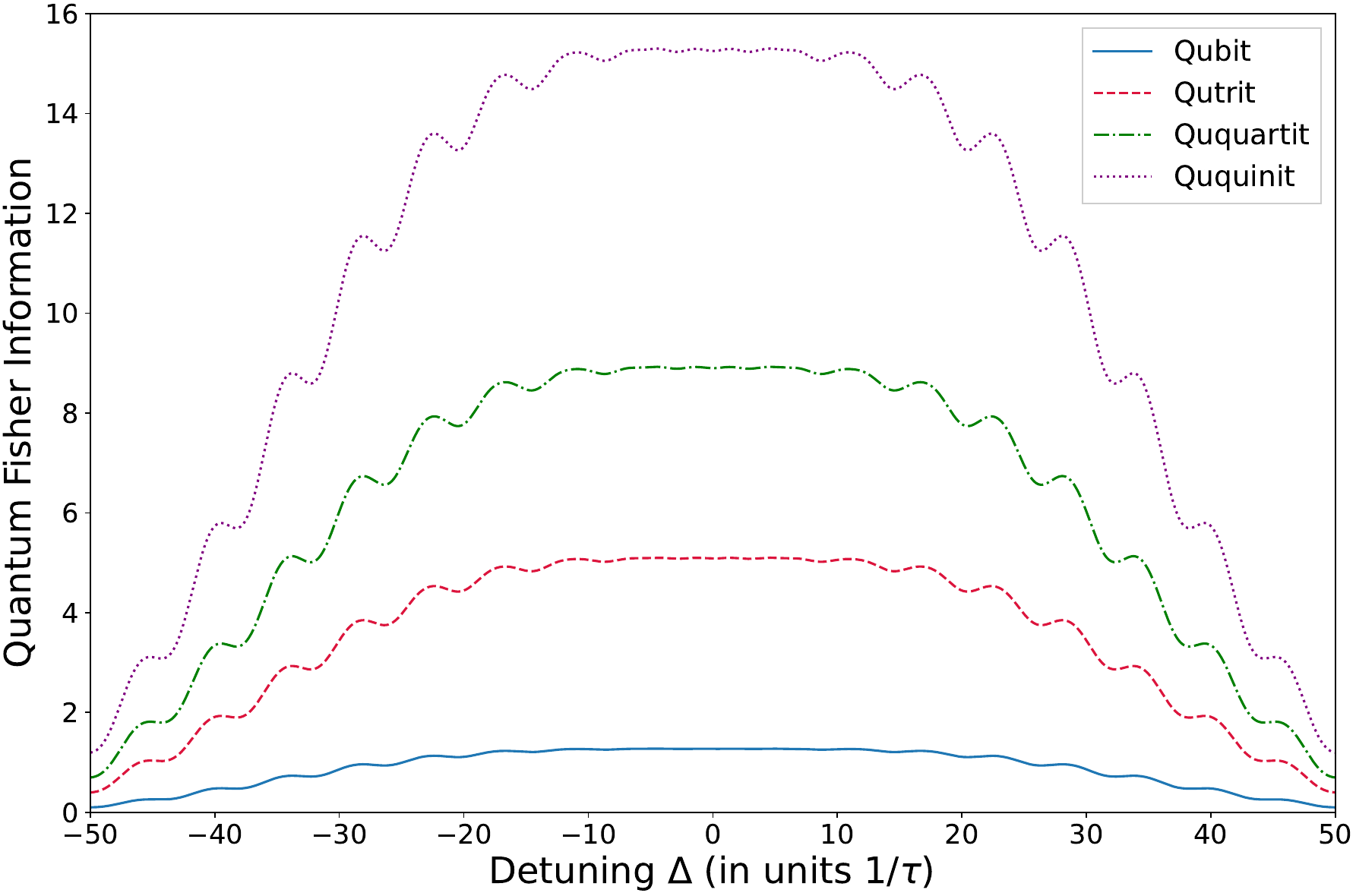}}
  \caption{Quantum Fisher information (QFI) for Ramsey interferometry versus detuning using Eq.~(\ref{eq:qfi_pure}).
  QFI is shown for qubits, qutrits, and selected higher-dimensional WM qudits. Other parameters are identical to those used in
  Fig.~\ref{fig:qft}.}
  \label{fig:qfi}
\end{figure}

Wigner--Majorana (WM) Ramsey interferometry is based on the effective
spin-\(j\) dynamics of a \(D=2j+1\) dimensional ladder driven by a single
near-resonant field. Equivalently, it can be realized in multiphoton
or Raman-type implementations that reproduce the same embedded
\(\mathrm{SU}(2)\) structure. The key point is that the multilevel
manifold supplies a genuine multipath interferometer while the control
requirements remain close to those of ordinary qubit Ramsey
interferometry: each interaction zone is driven by a single field. This single-field description presumes that the WM manifold is equally spaced, so that all $\Delta m=\pm1$ transitions are simultaneously resonant. Equivalently, the Rabi frequency $\Omega$ must exceed any static anharmonicity---such as a quadratic Zeeman or tensor light shift---that would otherwise detune the outer transitions and break the embedded SU(2) structure. This condition is least demanding for the qutrit and becomes progressively more stringent as $D$ grows.

The WM pulse with amplitude \(\Omega=\pi/(2T)\) is the direct analogue of
a qubit \(\pi/2\) pulse in the spin-\(j\) representation. It does not,
however, create an equal superposition over all qudit states. For
example, in a qutrit manifold, a resonant WM pulse applied to an outer
state \(|1\rangle\) or \(|3\rangle\), corresponding to \(m=-1\) or
\(m=1\), produces the population distribution \((1/4,1/2,1/4)\), while
the same pulse applied to the central state \(|2\rangle\), corresponding
to \(m=0\), produces \((1/2,0,1/2)\). The enhancement discussed below
therefore does not come from preparing a uniform qudit superposition. It
comes from the detuning-dependent multipath phases generated by the WM
ladder itself.

We now discuss the WM Ramsey protocols in the order used in the numerical
analysis: first the analytically transparent qutrit case, then higher
odd-dimensional manifolds, and finally higher even-dimensional
manifolds.

\subsection{Qutrit WM Ramsey interferometry}
\label{sec:qutrit_wm}



For $D=3$, the WM manifold is the spin-1 representation. In the basis
$\{|1\rangle,|2\rangle,|3\rangle\}$, corresponding to magnetic quantum
numbers $m=-1,0,1$, the rotating-frame Hamiltonian is
\begin{equation}
H_3 =
\begin{pmatrix}
-\Delta & \Omega/\sqrt{2} & 0 \\
\Omega/\sqrt{2} & 0 & \Omega/\sqrt{2} \\
0 & \Omega/\sqrt{2} & \Delta
\end{pmatrix}.
\end{equation}
Equivalently, \(H_3=\Omega J_x+\Delta J_z\), where $J_x$ and $J_z$ are
the spin-1 angular-momentum matrices in the basis ordered as
$m=-1,0,1$. The detuning $\Delta$ is the unknown frequency offset to be
estimated.

We use the repeated-pulse Ramsey convention in this subsection. The full
finite-pulse propagator is therefore
\begin{equation}
U_3^{(\mathrm{rep})}(\Delta)
=
R_3(T,\Delta)F_3(\tau,\Delta)R_3(T,\Delta),
\end{equation}
where \(R_3(T,\Delta)=\exp[-i(\Omega J_x+\Delta J_z)T]\) is the driven
WM pulse and \(F_3(\tau,\Delta)=\exp(-i\Delta\tau J_z)\) is the
dark-evolution propagator. In matrix form,
\begin{equation}
F_3(\tau,\Delta)
=
\operatorname{diag}
\left(
e^{i\Delta\tau},
1,
e^{-i\Delta\tau}
\right).
\end{equation}

The simple origin of the qutrit enhancement is clearest in the
ideal-pulse limit. In this limit the detuning during the short Ramsey
pulses is neglected, while the detuning during the dark time is retained.
The resonant WM pulse is calibrated as a spin-1 $\pi/2$ rotation, so
\(\Omega T=\pi/2\). Thus
\(R_3(T,0)=R_x^{(1)}(\pi/2)=\exp[-i(\pi/2)J_x]\). Explicitly,
\begin{equation}
R_x^{(1)}\left(\frac{\pi}{2}\right)
=
\begin{pmatrix}
1/2 & -i/\sqrt{2} & -1/2 \\
-i/\sqrt{2} & 0 & -i/\sqrt{2} \\
-1/2 & -i/\sqrt{2} & 1/2
\end{pmatrix}.
\end{equation}

We prepare the central WM state $|2\rangle$, corresponding to $m=0$.
After the first resonant WM $\pi/2$ pulse, the state becomes
\(R_x^{(1)}(\pi/2)|2\rangle=-(i/\sqrt{2})(|1\rangle+|3\rangle)\). Thus
the first pulse maps the central state entirely into a coherent
superposition of the two outer states. This is the key difference from a
generic qudit-gate superposition: the two populated arms have magnetic
quantum numbers $m=-1$ and $m=1$, and therefore acquire opposite phases
during the dark time.

After free evolution for time $\tau$, the state is
\[
F_3(\tau,\Delta)
R_x^{(1)}\left(\frac{\pi}{2}\right)|2\rangle
=
-\frac{i}{\sqrt{2}}
\left(
e^{i\Delta\tau}|1\rangle
+
e^{-i\Delta\tau}|3\rangle
\right).
\]
The relative phase between the two outer arms is therefore $2\Delta\tau$.
The second WM pulse recombines these two paths back onto the central
state. The central-state Ramsey amplitude is
\[
A_3(\Delta)
=
\langle 2|
R_x^{(1)}\left(\frac{\pi}{2}\right)
F_3(\tau,\Delta)
R_x^{(1)}\left(\frac{\pi}{2}\right)
|2\rangle .
\]
Using
\(\langle 2|R_x^{(1)}(\pi/2)|1\rangle=-i/\sqrt{2}\) and
\(\langle 2|R_x^{(1)}(\pi/2)|3\rangle=-i/\sqrt{2}\), one obtains
\(A_3(\Delta)=-(e^{i\Delta\tau}+e^{-i\Delta\tau})/2\). Hence
\(A_3(\Delta)=-\cos(\Delta\tau)\). The measured central-state return
probability is therefore \(P_3(\Delta)=|A_3(\Delta)|^2\), which gives the
ideal qutrit WM Ramsey signal
\begin{equation}
P_3(\Delta)
=
\cos^2(\Delta\tau).
\label{eq:qutrit_wm_signal}
\end{equation}

This should be compared with the ideal repeated-pulse qubit Ramsey signal
\(P_2(\Delta)=\cos^2(\Delta\tau/2)\). The qutrit signal therefore has
twice the central fringe density at the same interrogation time~\cite{ZhouLi2026}.
Equivalently, the maximal central-fringe slope is doubled in the
ideal-pulse limit: \(S_3=\tau\), while for the qubit
\(S_2=\tau/2\). Thus \(S_3/S_2=2\).

The physical origin of this factor of two is transparent from the path
picture above. In the qubit Ramsey sequence, the two interferometer arms
differ by one unit of the detuning phase, so the relevant relative phase is
$\Delta\tau$. In the spin-1 qutrit sequence, the first WM pulse maps the
central state onto the two outer states, whose energies differ by
$2\Delta$ in the rotating frame. Their relative phase is therefore
$2\Delta\tau$, and the recombined population oscillates as
$\cos^2(\Delta\tau)$. 

For finite rectangular pulses, the detuning also acts during the two
interaction zones. The exact pulse propagator can be written in terms of
the Cayley--Klein parameters $a$ and $b$ of the associated spin-1/2
problem as
\begin{equation}
R_3(T,\Delta)
=
\begin{pmatrix}
a^2 & \sqrt{2}ab & b^2 \\
-\sqrt{2}ab^\ast & |a|^2-|b|^2 & \sqrt{2}a^\ast b \\
b^{\ast 2} & -\sqrt{2}a^\ast b^\ast & a^{\ast 2}
\end{pmatrix}.
\end{equation}
Defining \(q=|a|^2-|b|^2\), the repeated-pulse central-state amplitude is
\begin{equation}
A_3^{(\mathrm{rep})}(\Delta)
=
q^2
-
2|b|^2
\left(
a^2 e^{i\Delta\tau}
+
a^{\ast 2} e^{-i\Delta\tau}
\right).
\end{equation}
The corresponding finite-pulse signal is
\(P_3^{(\mathrm{rep})}(\Delta)=|A_3^{(\mathrm{rep})}(\Delta)|^2\). In the
resonant ideal-pulse limit, $a=1/\sqrt{2}$, $b=-i/\sqrt{2}$, and $q=0$,
so this expression reduces to the simple result
$P_3(\Delta)=\cos^2(\Delta\tau)$.

\begin{figure}[!t]
  \centering
  \includegraphics[width=\linewidth]{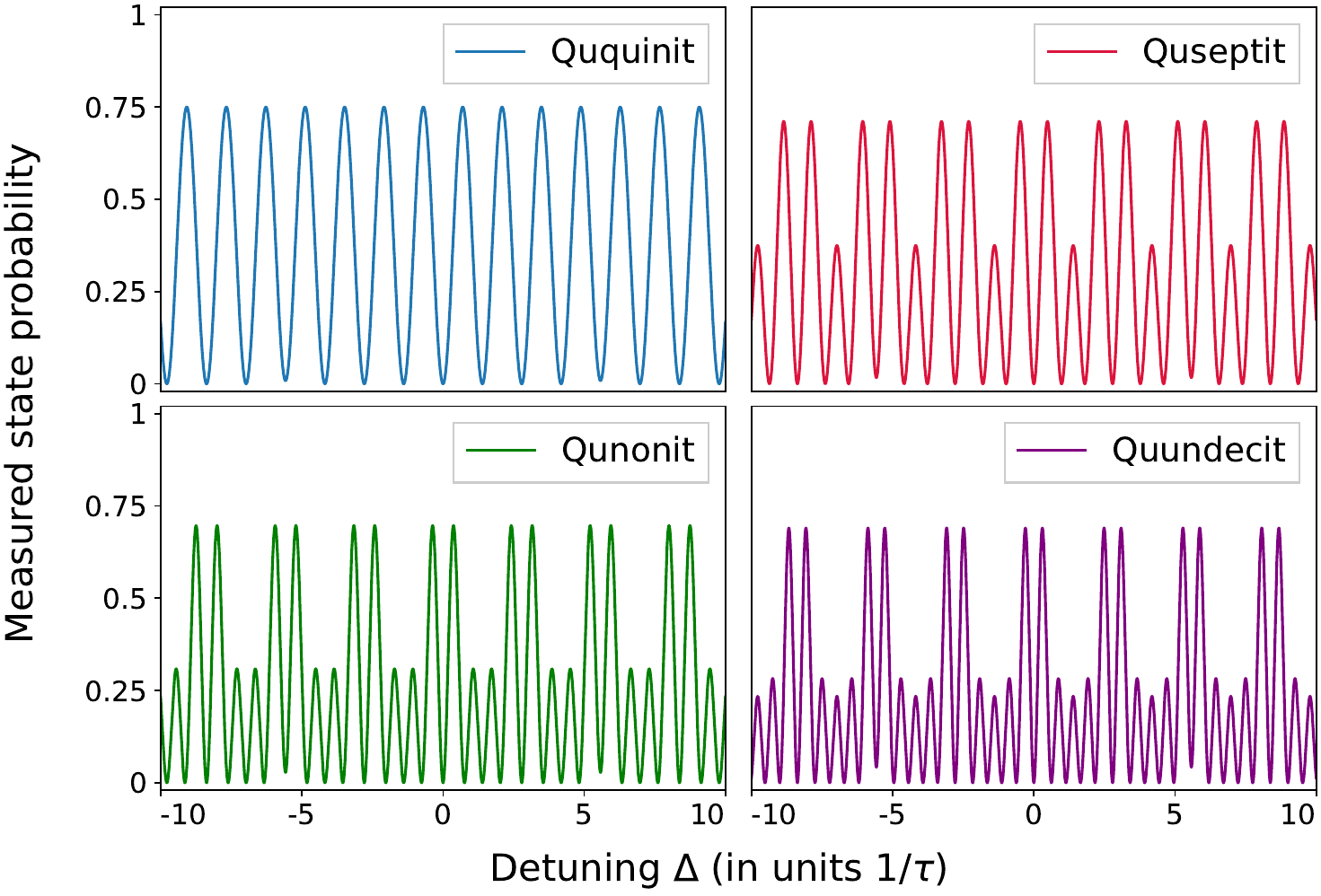}
  \caption{Comparison of odd-dimensional WM qudits with $D=5,7,9,11$. The plotted signals are the fixed shoulder-sum readouts defined in the text as functions of detuning. Other parameters are identical to those used in Fig.~\ref{fig:qft}.}
  \label{fig:quoddits}
\end{figure}

\begin{table}[!t]
\centering
\caption{Prepared states and scalar readouts used for the WM Ramsey protocols.}
\label{table:readouts}
\begin{tabular}{c c c}
\hline
$D$ & prepared state $|\ell\rangle$ & scalar readout $P_D(\Delta)$ \\
\hline
2 & $|1\rangle$ & $P_{1\rightarrow 2}(\Delta)$ \\
3 & $|2\rangle$ & $P_{2\rightarrow 2}(\Delta)$ \\
odd $D>3$ & $|(D+1)/2\rangle$ & $P_{\ell\rightarrow \ell-1}(\Delta)+P_{\ell\rightarrow \ell+1}(\Delta)$ \\
even $D>3$ & $|D/2\rangle$ & $P_{\ell\rightarrow \ell-1}(\Delta)+P_{\ell\rightarrow \ell+1}(\Delta)$ \\
\hline
\end{tabular}
\end{table}

\subsection{Odd-dimensional WM interrogations for \(D>3\)}
\label{sec:odd_wm}

For odd \(D\), the WM manifold contains a unique central state with
\(m=0\). In the chosen basis, this state has index \(\ell=(D+1)/2\). The
natural initial state is therefore the central state \(|\ell\rangle\).
Unlike the qutrit case, however, the dominant near-central oscillatory
response for \(D>3\) is not captured most effectively by a single return
probability. Inspection of the full WM Ramsey propagators shows that the
signal is distributed mainly over the two symmetric nearest-neighbor
channels adjacent to the prepared state. We therefore use a binned
shoulder readout. The prepared states and scalar readouts used across all dimensions are summarized in Table~\ref{table:readouts}.

The measured set of output channels is \(I_D=\{\ell-1,\ell+1\}\).
This fixed rule is used for all \(D>3\) because the return channel is no longer the dominant high-slope central feature; the nearest-neighbor shoulder channels carry the largest useful near-central oscillatory response in the population-transfer plots shown in Appendix~\ref{appendix_a}. This gives an experimentally simple and operationally consistent comparison across dimensions. Equivalently, the readout projector is
\begin{equation}
\Pi_D =
|\ell-1\rangle\langle \ell-1|
+
|\ell+1\rangle\langle \ell+1|.
\end{equation}
The odd-\(D\) WM Ramsey signal is then defined as the corresponding
shoulder-sum probability,
\begin{equation}
P_D(\Delta)
=
P_{\ell\rightarrow \ell-1}(\Delta)
+
P_{\ell\rightarrow \ell+1}(\Delta),
\qquad
\ell=\frac{D+1}{2}.
\end{equation}

For \(D=5\), this gives
\begin{equation}
P_5(\Delta)
=
P_{3\rightarrow 2}(\Delta)
+
P_{3\rightarrow 4}(\Delta).
\end{equation}
The same construction applies to the higher odd dimensions studied
numerically, such as \(D=7,9,11\). The resulting traces show an
increasing density of near-central Ramsey features as \(D\) grows. 

Across the odd-dimensional data set, the extracted central-fringe slopes
increase with \(D\), but the contrast does not remain constant. Instead,
the higher-dimensional enhancement comes with a visibility cost. In the
parameter regime studied here, the $D=5$ WM qudit, still offers a
favorable balance between increased slope and usable contrast, while
larger odd dimensions continue to sharpen the fringes at the cost of
reduced visibility.

This behavior illustrates the main resolution--contrast trade-off of WM
qudit Ramsey interferometry. The larger effective separation between
interfering WM pathways increases the detuning response, but the
population signal becomes distributed over more channels. A binned
readout recovers the dominant shoulder response, yet the central-fringe
contrast generally decreases with dimension. 

\begin{figure}[!t]
  \centering
  \includegraphics[width=\linewidth]{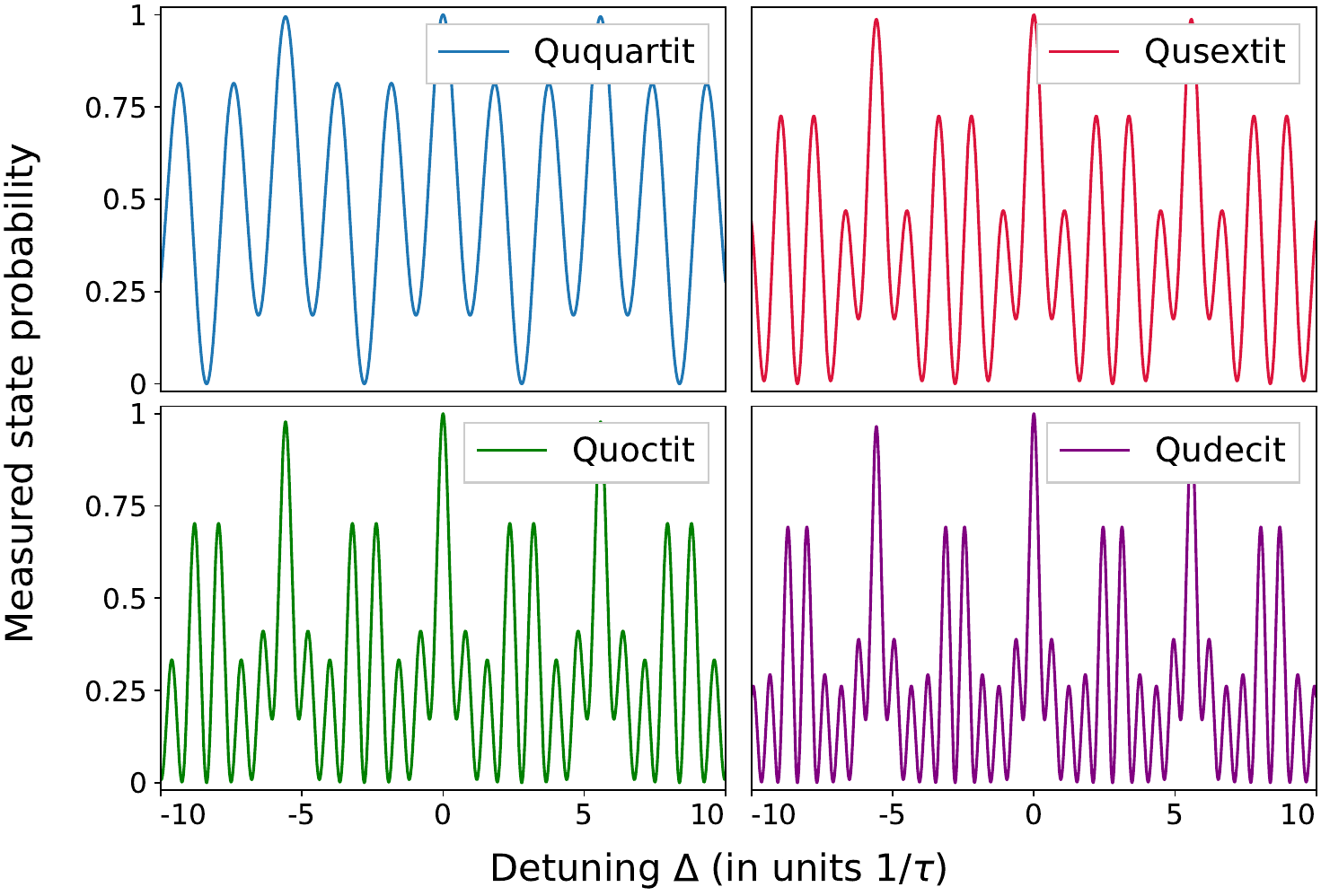}
  \caption{Comparison of even-dimensional WM qudits with $D=4,6,8,10$.  The plotted signals are the fixed shoulder-sum readouts defined in the text as functions of detuning. Other parameters are identical to those used in Fig.~\ref{fig:qft}.}
  \label{fig:qunoddits}
\end{figure}

\begin{figure}[!t]
  \centering
  \includegraphics[width=\linewidth]{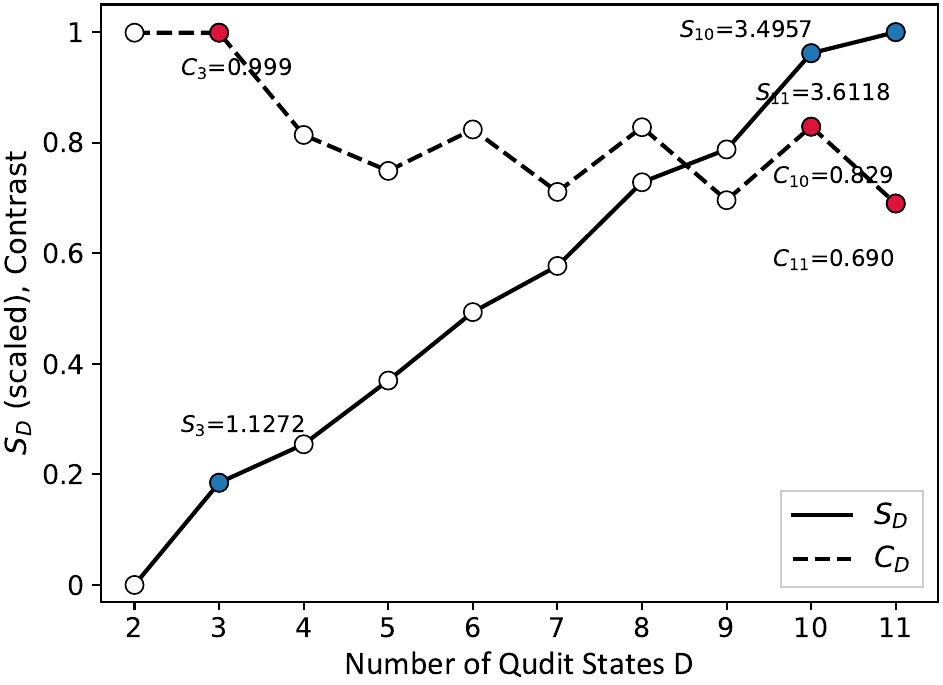}
  \caption{Resolution--contrast trade-off for WM Ramsey readouts. The solid line shows the central-fringe slope $S_D$ rescaled for visual comparison, while the dashed line shows the corresponding contrast $C_D$. The scatter points represent the simulated protocols with $D=2,\ldots,11$, with annotations for selected dimensions.}
  \label{fig:pareto}
\end{figure}

\subsection{Even-dimensional WM interrogations for \(D>3\)}
\label{sec:even_wm}

For even \(D\), the WM manifold has no unique \(m=0\) state. Instead,
there are two symmetry-related central states with \(m=\pm 1/2\). In the
chosen basis these correspond to the indices \(\ell=D/2\) and
\(\ell+1=D/2+1\). These two states yield equivalent Ramsey responses by
symmetry. Without loss of generality, we prepare the lower of the two
central states, \(|\ell\rangle\).

As for odd \(D>3\), a single output population does not capture the full
dominant oscillatory response. The strongest near-central contribution
is again shared by two symmetric nearest-neighbor shoulder channels
adjacent to the prepared state. We therefore use the binned readout
\(I_D=\{\ell-1,\ell+1\}\). The corresponding readout projector is
\begin{equation}
\Pi_D =
|\ell-1\rangle\langle \ell-1|
+
|\ell+1\rangle\langle \ell+1|.
\end{equation}
The even-\(D\) WM Ramsey signal is defined as
\begin{equation}
P_D(\Delta)
=
P_{\ell\rightarrow \ell-1}(\Delta)
+
P_{\ell\rightarrow \ell+1}(\Delta),
\qquad
\ell=\frac{D}{2}.
\end{equation}

For \(D=4\), this becomes
\begin{equation}
P_4(\Delta)
=
P_{2\rightarrow 1}(\Delta)
+
P_{2\rightarrow 3}(\Delta).
\end{equation}
The same construction is used for the even-dimensional manifolds
\(D=6,8,10\). The corresponding Ramsey traces exhibit a dominant
central feature together with smaller satellite oscillations. As in the
odd-dimensional case, the central-fringe slope increases with dimension,
while the contrast is reduced relative to the qutrit.

The even-dimensional protocols therefore follow the same qualitative
trend as the odd-dimensional ones: higher \(D\) gives stronger fringe
compression and larger central slopes, but the usable contrast becomes
dimension dependent. In the numerical data, the largest even dimensions
studied provide some of the steepest central slopes, but their contrast
is lower than the qutrit contrast. Thus they are useful when maximal
local slope is the priority, whereas the qutrit remains the most
balanced operating point when both slope and contrast are considered.

For completeness, Fig.~\ref{fig:qfi} reports the quantum Fisher information (QFI)~\cite{BraunsteinCaves1994,Paris2008} of the detuning-dependent output states. For a pure state $|\psi(\Delta)\rangle$, we use the standard pure-state form~\cite{Fujiwara1995,AnandanAharonov1990}
\begin{equation}
F_Q(\Delta)
=
4\left(
\langle \partial_\Delta\psi|\partial_\Delta\psi\rangle
-
|\langle \psi|\partial_\Delta\psi\rangle|^2
\right).
\label{eq:qfi_pure}
\end{equation}
The QFI provides a measurement-independent benchmark~\cite{olivares2025quantum}, while the fixed population readouts used experimentally are compared through $S_D$, $C_D$, and the classical Fisher information defined in Appendix~\ref{app:ramsey-metrics}. For the ideal qutrit binary readout, $P_3(\Delta)=\cos^2(\Delta\tau)$ gives $F_{\mathrm{cl},3}=4\tau^2$ and therefore $\delta\Delta_3^{\mathrm{opt}}=1/(2\tau\sqrt{N})$, a factor-of-two improvement over the ideal qubit baseline.


\begin{table}[!t]
  \centering
  \scriptsize
  \setlength{\tabcolsep}{3pt}
  \renewcommand{\arraystretch}{1.25}
  \begin{tabular}{||c|c|c|c|c||}
    \hline
    \rule{0pt}{2.4ex}$D$ & $S_D$ & $C_D$ & $S_D/S_2$ & $\mathcal{F}_D$ \\[0.4ex]
    \hline\hline
    2  & 0.5635 & 0.999 & 1.00 & 1.00 \\
    \hline
    3  & 1.1272 & 0.999 & 2.00 & 4.00 \\
    \hline
    4  & 1.3392 & 0.814 & 2.38 & 5.65 \\
    \hline
    5  & 1.6908 & 0.749 & 3.00 & 9.00 \\
    \hline
    6  & 2.0680 & 0.824 & 3.67 & 13.47 \\
    \hline
    7  & 2.3226 & 0.711 & 4.12 & 16.99 \\
    \hline
    8  & 2.7840 & 0.828 & 4.94 & 24.41 \\
    \hline
    9  & 2.9651 & 0.696 & 5.26 & 27.69 \\
    \hline
    10 & 3.4957 & 0.829 & 6.20 & 38.48 \\
    \hline
    11 & 3.6118 & 0.690 & 6.41 & 41.08 \\
    \hline
  \end{tabular}
  \caption{Central-fringe metrics extracted from simulations. Here $S_D$ is the maximal absolute slope of the central fringe [Eq.~(\ref{eq:SD})] under the fixed scalar readout $P_D(\Delta)$ [Eq.~\eqref{eq:PD}], and $C_D$ is the corresponding contrast [Eq.~(\ref{eq:CD})]. The final column is the slope-squared sensitivity proxy $\mathcal{F}_D=(S_D/S_2)^2$, which equals the normalized classical Fisher information for an ideal unit-contrast binary Ramsey fringe; for reduced-contrast traces the full $F_{\mathrm{cl},D}$ should be evaluated from Appendix~\ref{app:ramsey-metrics}. The slopes are extracted from the finite-pulse simulations with $\tau=1$ and $T=0.1$, and therefore differ slightly from the ideal-pulse values.}
  \label{table:1}
\end{table}

\section{Decoherence effects in Wigner--Majorana systems}
\label{sec:decoherence}

Ramsey interferometry is directly sensitive to decoherence, because the
detuning information is encoded in the relative phases accumulated by
the interferometer pathways. In a standard two-level setting, the
relevant timescales are the population relaxation time \(T_1\), the
coherence time \(T_2\), and the pure dephasing time \(T_\phi\), with
\(1/T_2 = 1/(2T_1)+1/T_\phi\).

In the present work we focus on probe-induced phase noise rather than
population relaxation. This is motivated by precision-spectroscopy
settings in which a dominant systematic effect is a fluctuating
differential light shift, or probe shift~\cite{Yudin2023EllipticityShift,AbdelHafiz2022SABR}, associated with the probe fields
used for the Ramsey interaction. Such fluctuations imprint random
phases on the interferometer arms and therefore reduce the contrast of
the Ramsey fringes.

We model this effect using a Markovian GKSL master equation~\cite{Manzano2020} for the
density operator \(\rho_D(t)\),
\begin{align}
\dot{\rho}_D(t)
&=
-i[H_D,\rho_D(t)] \notag \\
&+
\sum_k \Gamma_k
\left(
L_k\rho_D(t)L_k^\dagger
-
\frac{1}{2}
\{L_k^\dagger L_k,\rho_D(t)\}
\right),
\label{eq:lindblad}
\end{align}
where \(H_D\) is the rotating-frame WM Hamiltonian, \(L_k\) are jump
operators, and \(\Gamma_k\) are the corresponding rates. To isolate the
effect of phase noise, we set population relaxation to zero,
\(\Gamma_1\simeq 0\). The pure-dephasing rate is parametrized as
\(\Gamma_\phi=2/T_\phi\). In the simulations shown below, this
Lindblad term is included over the full Ramsey sequence, i.e. during
both driven Ramsey zones and the dark interval, so that \(T_\phi\)
acts as a single effective phase-noise timescale for the complete
interrogation. If a specific platform has probe-shift fluctuations only
during part of the sequence, the same equations apply after replacing
\(\Gamma_\phi t\) by the appropriate integrated dephasing strength.

The probe-shift noise considered here is represented by a single
diagonal jump operator that projects onto the initially prepared state,
\(L_\phi=|\ell\rangle\langle \ell|\). Here \(|\ell\rangle\) is the
prepared WM state used in the corresponding Ramsey protocol. Physically,
this describes fluctuations of a diagonal energy shift that distinguish
the prepared basis states from the remaining WM manifold. Coherences involving
\(|\ell\rangle\) therefore acquire random phase diffusion, while
coherences entirely within the complementary subspace are unaffected by
this particular common-mode noise channel.

For the projector noise model, each density-matrix element is an
eigenmode of the dissipator. The dephasing contribution is
\begin{equation}
\left.\dot{\rho}_{mn}\right|_\phi
=
-\frac{\Gamma_\phi}{2}
|\delta_{m\ell}-\delta_{n\ell}|^2
\rho_{mn}.
\end{equation}
Thus populations are unchanged, and only coherences connecting the
prepared state to the rest of the manifold decay. For \(n\neq \ell\),
\(\rho_{\ell n}(t)\propto e^{-(\Gamma_\phi/2)t}\). Likewise,
\(\rho_{n\ell}(t)\propto e^{-(\Gamma_\phi/2)t}\). By contrast,
coherences \(\rho_{mn}\) with \(m,n\neq \ell\) are not damped by this
channel.

It is useful to contrast this projector model with a general diagonal
jump operator \(L=\sum_m \lambda_m |m\rangle\langle m|\). In that case
the corresponding dephasing of each coherence is
\begin{equation}
\left.\dot{\rho}_{mn}\right|_\phi
=
-\frac{\Gamma_\phi}{2}
(\lambda_m-\lambda_n)^2
\rho_{mn}.
\end{equation}
The projector model is therefore a special diagonal-noise channel in
which only the prepared arm is distinguished from the rest of the WM
manifold. For \(D>3\), this projector should not be identified with the
most general tensor-light-shift or quadratic-Zeeman noise operator. A
literal quadratic shift would normally be modeled by a diagonal operator
such as \(J_z^2\), which is equivalent to the projector model only in
the spin-\(1\) qutrit case discussed below, up to an irrelevant identity
shift. The higher-dimensional simulations therefore test a minimal
arm-selective common-mode dephasing model, not universal robustness to
all even-in-\(m\) diagonal shifts.

\subsection{Qutrit dephasing}

\begin{figure}[!t]
  \centerline{\includegraphics[width=\linewidth]{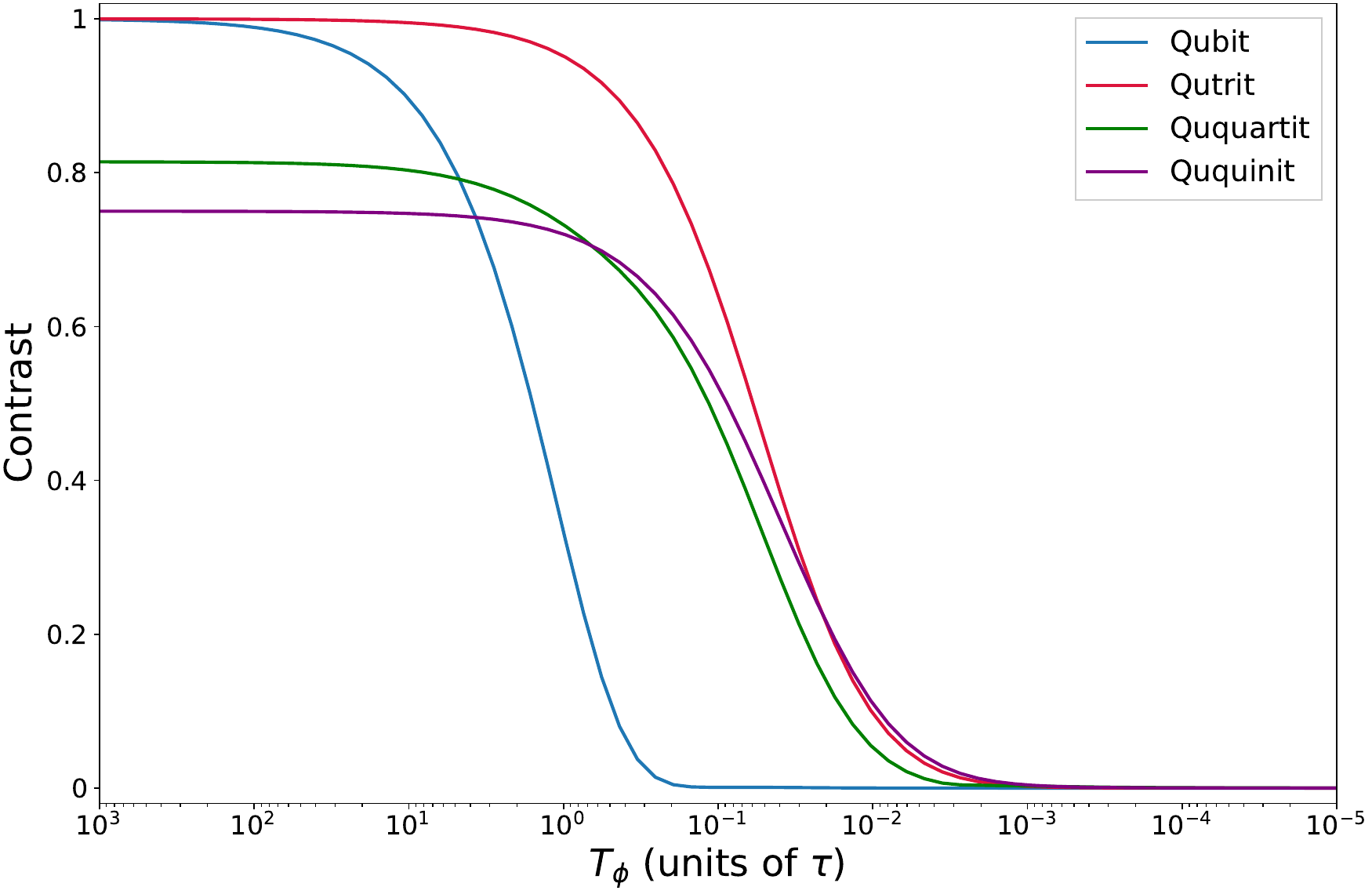}}
  \caption{Central fringe contrast for a qubit, qutrit, $D=4$ and $D=5$ WM qudits under probe-shift-induced
  dephasing $T_\phi$ only ($\Gamma_1\approx 0$) following Eq.~(\ref{eq:lindblad}). Within the common-mode
  probe-shift model described in the text ($L_\phi=|\ell\rangle\langle \ell|$), the selected qudit fringes remain visible over a broader dephasing range than the corresponding qubit readout for the parameter set shown here. The horizontal $T_\phi$ axis is logarithmic and given in units of the interrogation time $\tau$, and decreases from left to right, so that the dephasing strength increases toward the right.
  All other parameters are identical to those used in Fig.~\ref{fig:qft}.}
  \label{fig:ramsey_decoherence}
\end{figure}

For the qutrit protocol discussed in Sec.~\ref{sec:qutrit_wm}, the initial state is the
central WM state, \(|\ell\rangle=|2\rangle\). The dephasing jump operator
is therefore \(L_\phi=|2\rangle\langle 2|\). Using the result above, the
coherences connecting the central state to the two outer states decay as
\(\rho_{21}(t)\propto e^{-(\Gamma_\phi/2)t}\) and
\(\rho_{23}(t)\propto e^{-(\Gamma_\phi/2)t}\). The outer-state coherence
\(\rho_{13}\), however, is unaffected by this particular channel.

This structure has a simple symmetry interpretation. In the spin-\(1\)
WM manifold, the two outer states correspond to \(m=\pm 1\), while the
central state corresponds to \(m=0\). The projector noise model describes
an even-in-\(m\), common-mode diagonal shift of the two outer states
relative to the central state. Up to an identity shift, the same
dephasing channel can be represented by \(L_\phi=J_z^2\). Indeed, for
\(j=1\), \(J_z^2=I-|2\rangle\langle 2|\). Adding a multiple of the
identity to a Lindblad jump operator does not change the corresponding
pure-dephasing dissipator, so the projector model and the \(J_z^2\)-type
model describe the same dephasing structure in this case.

This is the relevant symmetry class for quadratic or even-in-\(m\)
shifts, such as tensor light shifts~\cite{DeutschJessen2010,Geremia2006}, quadratic Zeeman shifts, or
zero-field-splitting-type terms in spin-\(1\) platforms. For example, in
NV centers the ground-state Hamiltonian contains a term proportional to
\(S_z^2\)~\cite{Cambria2023ZFS,Hart2021DQ4Ramsey}. Fluctuations of this term dephase the \(m_s=0\) arm relative
to the \(m_s=\pm 1\) manifold while leaving the two outer states
common-mode.

The scalar part of the light shift completes this picture. A
fluctuating diagonal shift decomposes into scalar (\(m^0\)), vector
(\(m^1\)), and tensor (\(m^2\)) components~\cite{DeutschJessen2010,Geremia2006}. The scalar component shifts
every WM sublevel equally; as a jump operator it is proportional to the
identity, \(L_\phi\propto I\), and adding a multiple of the identity leaves
the pure-dephasing dissipator unchanged. Every coherence in the manifold
is therefore rigorously immune to scalar (common-mode intensity) noise,
independently of dimension --- an even more universal protection than the
even-in-\(m\) immunity of the outer coherence \(\rho_{13}\) described
above. Since the scalar AC Stark shift often dominates the light-shift
budget, this leaves only the vector (linear-in-\(m\)) component as a
relevant diagonal light-shift channel reaching \(\rho_{13}\); it carries
the same \(J_z\) structure as the linear Zeeman noise analyzed below~\cite{Yudin2023EllipticityShift}.

The numerical results show that, under this probe-shift-motivated
common-mode dephasing model, the qutrit Ramsey signal is less
contrast-sensitive over the plotted range of \(T_\phi\) than the corresponding
qubit signal. This is the robustness behavior shown in Fig.~\ref{fig:ramsey_decoherence}.

\subsection{Odd-dimensional qudit dephasing}

For higher odd-dimensional WM manifolds, \(D>3\), the prepared state is
again the unique central state, \(\ell=(D+1)/2\). The dephasing jump
operator is chosen as \(L_\phi=|\ell\rangle\langle \ell|\). This choice
gives the same per-coherence decay law for all odd dimensions:
coherences connecting the prepared central state to the remaining
manifold decay with rate \(\Gamma_\phi/2\), while coherences not involving
\(|\ell\rangle\) are unaffected by this particular projector channel.

The readout used for odd \(D>3\) is the shoulder-sum probability defined
in Sec.~\ref{sec:odd_wm}. Hence the relevant signal is dominated by coherences that
connect the prepared central state to the two neighboring shoulder
states. These coherences experience the calibrated decay envelope
\(e^{-(\Gamma_\phi/2)t}\). The comparison in Fig.~\ref{fig:ramsey_decoherence} indicates that the
odd-dimensional qudit signals can remain visible over a broader range of
probe-shift dephasing strengths than the corresponding qubit contrast
under the same scalar-readout convention.

The improvement should not be interpreted as universal immunity to
dephasing. It is a consequence of the specific common-mode diagonal-noise
structure assumed here. If the dominant noise instead distinguishes WM
states according to their magnetic quantum number, the behavior changes,
as discussed below.

\subsection{Even-dimensional qudit dephasing}

For even-dimensional WM manifolds, there is no unique central state.
Instead, the two central states have \(m=\pm 1/2\). In the convention of
Sec.~\ref{sec:even_wm}, we prepare the lower central state, \(\ell=D/2\). The same
projector dephasing model is then used, \(L_\phi=|\ell\rangle\langle
\ell|\). The relevant coherences connecting the prepared state to the two
nearest-neighbor shoulder channels decay with the same calibrated
envelope as in the odd-dimensional case, \(e^{-(\Gamma_\phi/2)t}\). The
even-dimensional signals therefore exhibit the same qualitative behavior:
the WM Ramsey fringes can retain appreciable contrast under the
common-mode probe-shift noise model, although the absolute contrast also
depends on the dimension-dependent readout contrast already present in
the coherent dynamics.

\subsection{Comparison with linear Zeeman dephasing}

The robustness described above is specific to the projector-type,
common-mode probe-shift model. A standard alternative is linear Zeeman
dephasing, represented by \(L=J_z\). In the general diagonal-noise formula
this corresponds to \(\lambda_m=m\). The decay rate of a coherence
\(\rho_{mn}\) is then proportional to \((m-n)^2\).

For the WM Ramsey readouts used here, however, the compressed central
fringe is not carried by adjacent coherences but by the \emph{outer}
coherence between the extreme WM states \(m=\pm(D-1)/2\) --- the same
coherence \(\rho_{13}\) that is left invariant by the even-in-\(m\)
projector channel above and that produces the fringe sharpening in the
first place. This coherence has \(|m-n|=D-1\), so under \(L=J_z\) it
dephases at the enhanced rate \((\Gamma_\phi/2)(D-1)^2\). The qudit is
therefore more sensitive to linear Zeeman noise than the qubit, by
a factor that grows with dimension. Full-sequence simulations give
half-contrast coherence times in the ratio \(1:4:8:16\) for
\(D=2,3,4,5\), so the qutrit already requires about four times longer
coherence to reach the same contrast. Robustness to common-mode probe
shifts and sensitivity to linear Zeeman noise are thus two sides of the
same outer coherence---the fringe compression that sharpens the resolution
is precisely what amplifies the linear-Zeeman dephasing.

The main conclusion is therefore twofold. First, WM qudit Ramsey
interferometry can be robust against experimentally relevant
common-mode, even-in-\(m\) probe-shift fluctuations. Second, this
robustness is not a general property of all diagonal dephasing channels.
It depends on the symmetry of the dominant noise source relative to the
WM manifold and to the chosen scalar readout.


\section{Discussion}
\label{sec:discussion}

We have developed a single-qudit extension of Ramsey interferometry based
on Wigner--Majorana (WM) spin dynamics. The central idea is to use the
internal multilevel structure of a single quantum system as a multipath
interferometer. In this setting, the qudit dimension \(D\) is not merely a
larger Hilbert space for information encoding, but an interferometric
resource that modifies the detuning response at fixed interrogation time
\(\tau\).

All dimensions were compared using the same operational framework. For
each \(D\), the Ramsey output was reduced to a scalar population readout
\(P_D(\Delta)\), defined either as a single measured channel or as a
binned sum of several channels. The central Ramsey feature was then
characterized by two experimentally meaningful quantities: the maximal
central-fringe slope \(S_D\) and the corresponding contrast \(C_D\). This
comparison is important because a narrow or steep feature is useful only
if it retains sufficient contrast for reliable readout.

The clearest enhancement occurs for the qutrit. In the WM qutrit
protocol, the system is prepared in the central state of the spin-\(1\)
manifold and the central-state return probability is measured. In the
ideal-pulse limit, the qutrit signal is
\begin{equation*}
P_3(\Delta) = \cos^2(\Delta \tau).
\end{equation*}
This should be compared with the qubit Ramsey signal
\begin{equation*}
P_2(\Delta) = \cos^2\left(\frac{\Delta \tau}{2}\right).
\end{equation*}
Thus the qutrit central fringe is compressed by a factor of
two relative to the qubit fringe at the same interrogation time in the ideal limit. The
numerical data confirm this result: the extracted slope increases from
\(S_2 \simeq 0.564\) to \(S_3 \simeq 1.127\), while the contrast remains
essentially unchanged, \(C_2 \simeq C_3 \simeq 1\). The qutrit therefore
emerges as the most balanced operating point in the present study,
providing a substantial resolution gain without a visibility penalty.

For higher odd-dimensional WM manifolds, \(D>3\), the system again has a
unique central state. However, the dominant near-central oscillatory
response is no longer best captured by the return probability alone.
Instead, it is mainly distributed over the two symmetric nearest-neighbor
shoulder channels adjacent to the prepared state. For this reason, we use
the binned readout
\begin{equation*}
P_D(\Delta)
=
P_{\ell\rightarrow \ell-1}(\Delta)
+
P_{\ell\rightarrow \ell+1}(\Delta),
\qquad
\ell=\frac{D+1}{2}.
\end{equation*}
The odd-dimensional data show that the central-fringe slope continues to
increase with \(D\). At the same time, the contrast becomes
dimension-dependent and generally decreases compared with the qutrit
case. The $D=5$ WM qudit still offers a favorable trade-off in the
parameter regime studied here, but larger odd dimensions increasingly
illustrate the general resolution--contrast compromise: the fringes
become sharper, yet the usable signal amplitude is reduced.

Even-dimensional WM manifolds behave similarly, but with a different
central-state structure. Since there is no unique \(m=0\) state, the two
central states correspond to \(m=\pm 1/2\). We prepare one of these two
states, chosen as
\begin{equation*}
\ell=\frac{D}{2},
\end{equation*}
and again use the nearest-neighbor shoulder-sum readout
\begin{equation*}
P_D(\Delta)
=
P_{\ell\rightarrow \ell-1}(\Delta)
+
P_{\ell\rightarrow \ell+1}(\Delta).
\end{equation*}
For the even dimensions studied here, \(D=4,6,8,10\), the Ramsey traces
show a dominant central feature together with satellite oscillations. The
largest even dimensions provide some of the steepest central slopes in
the numerical data, but their contrast is lower than the qutrit contrast.
They are therefore attractive when maximal local slope is the primary
objective, while the qutrit remains preferable when one wants a robust
combination of high slope and high contrast.

This behavior summarizes the main trade-off of WM qudit Ramsey
interferometry. Increasing \(D\) increases the density of the central
features and hence the local detuning response. However, the same
multilevel interference that sharpens the signal also redistributes
population over more output channels, which can reduce the contrast of
any fixed scalar readout. The optimal dimension therefore depends on the
experimental priority. If the goal is the best overall balance between
slope and contrast, the qutrit is the natural choice. If the goal is
maximum local fringe compression, higher-dimensional qudits may be
advantageous, provided that the reduced contrast can be tolerated.

The present comparison deliberately uses a fixed and experimentally simple
population readout. Optimized binnings or weighted linear combinations of
output populations could improve the contrast and Fisher information of
the higher-dimensional protocols, but such optimized readouts are left for
future work.

An important practical point is that the WM enhancement does not require
the control complexity usually associated with general qudit gates. Each
Ramsey interaction zone is generated by a single near-resonant field
driving the WM ladder. The multistate manifold supplies the additional
interferometer pathways internally. This distinguishes the WM protocol
from gate-based qudit Ramsey variants based on \(\mathrm{QFT}_D\) or
\(\sqrt{X_D}\) operations. In the ideal-gate model used here, those gates
are treated as \(\Delta\)-independent, so the detuning enters only during
free evolution. As a result, they do not provide the same regular
central-fringe compression as the WM dynamics under the scalar readouts
considered in this work.

The decoherence analysis further clarifies when the WM advantage is
expected to survive in realistic settings. We focused on a
probe-shift-motivated diagonal dephasing model in which the prepared arm
is dephased relative to the remaining WM manifold. In the qutrit case,
this common-mode projector model is equivalent, up to an identity shift,
to an even-in-\(m\) dephasing structure such as that produced by tensor
light shifts, quadratic Zeeman shifts, or zero-field-splitting-type
fluctuations in spin-\(1\) systems. For higher dimensions, we use the
projector as a minimal arm-selective model rather than as a generic model
of all quadratic shifts. Under this noise model, the selected qudit
fringes can remain visible in regimes where the qubit signal is strongly
contrast-suppressed under the same readout convention.

This robustness should not be interpreted as a universal immunity of WM
interferometry to dephasing. For linear Zeeman noise, \(L=J_z\), the
same large \(m\)-separation that produces fringe compression also enhances
dephasing. The dominant compressed-fringe coherence has \(|m-n|=D-1\), so
its dephasing rate scales as \((D-1)^2\). Thus higher-dimensional WM
readouts are not only non-protected against \(J_z\) noise; they can become
increasingly sensitive to it. The noise benefit is therefore
symmetry-selective: it depends on how the dominant physical noise channel
acts on the WM manifold.

Overall, the results identify WM qudits as a practical route to enhanced
Ramsey spectroscopy at fixed interrogation time. The qutrit provides the
most immediate gain, doubling the central-fringe slope in the ideal limit
without sacrificing contrast. Higher odd and even dimensions can further
increase the slope, but at the price of a dimension-dependent contrast
reduction. Future improvements may come from optimizing the scalar
readout, combining WM interrogation with composite or hyper-Ramsey pulse
designs~\cite{Wimperis1994,Genov2011,Beloy2018,vitanov2015fault}, and tailoring the protocol to the dominant noise symmetry of a
specific experimental platform.

\section{Conclusion}
\label{sec:conclusion}

We have introduced a single-qudit extension of Ramsey interferometry based
on Wigner--Majorana (WM) spin dynamics. In this approach, the internal
multilevel structure of a single quantum system acts as a multipath
interferometer. The enhancement does not rely on multipartite
entanglement, squeezed input states~\cite{Wineland1992,Giovannetti2011}, or longer interrogation times.
Instead, it arises from the detuning-dependent interference of pathways
inside a driven WM ladder.

The most transparent result is obtained for the qutrit. When the system
is prepared in the central state of the spin-\(1\) manifold, the WM
Ramsey signal has a doubled central-fringe density
relative to the qubit signal at the same dark time \(\tau\) in the
ideal-pulse limit. The qutrit response is
\begin{equation*}
P_3(\Delta)
=
\cos^2(\Delta \tau),
\end{equation*}
whereas the corresponding qubit Ramsey response is
\begin{equation*}
P_2(\Delta)
=
\cos^2\left(\frac{\Delta \tau}{2}\right).
\end{equation*}
The numerical simulations confirm this doubling at the level of the
central-fringe slope: \(S_3\) is approximately twice \(S_2\), while the
contrast remains essentially unity. Thus the qutrit provides the most
favorable operating point in the parameter regime studied here, combining
a clear resolution gain with high-visibility fringes.

Higher odd-dimensional WM manifolds further increase the central-fringe
density. For \(D>3\), the useful signal is not captured most effectively
by the central-state return probability alone. Instead, the dominant
near-central response is distributed over the two symmetric
nearest-neighbor shoulder channels, motivating the binned readout
\begin{equation*}
P_D(\Delta)
=
P_{\ell\rightarrow \ell-1}(\Delta)
+
P_{\ell\rightarrow \ell+1}(\Delta),
\qquad
\ell=\frac{D+1}{2}.
\end{equation*}
Within this readout, the odd-dimensional protocols show increasing slope
with increasing \(D\), but also a dimension-dependent reduction in
contrast. This identifies a general resolution--contrast trade-off:
larger odd-dimensional qudits can sharpen the detuning response, but the
usable readout visibility is reduced.

Even-dimensional WM manifolds exhibit the same qualitative behavior, with
the additional feature that there is no unique central \(m=0\) state.
Instead, the two central states correspond to \(m=\pm 1/2\). Preparing one
of these states and using the same shoulder-sum readout,
\begin{equation*}
P_D(\Delta)
=
P_{\ell\rightarrow \ell-1}(\Delta)
+
P_{\ell\rightarrow \ell+1}(\Delta),
\qquad
\ell=\frac{D}{2},
\end{equation*}
also produces compressed central Ramsey features. The largest even
dimensions studied here yield some of the steepest slopes, but with
lower contrast than the qutrit. They are therefore useful when maximal
local slope is prioritized, whereas the qutrit remains the best balanced
choice when both slope and contrast are important.

A central practical advantage of the WM construction is that it does not
require the implementation of general qudit gates such as \(\mathrm{QFT}_D\) or
\(\sqrt{X_D}\). Each Ramsey interaction zone is generated by a single
near-resonant drive acting on the WM ladder. The additional
interferometric pathways are supplied by the internal structure of the
qudit itself. This keeps the control workload close to that of standard
qubit Ramsey interferometry while enabling a stronger detuning response.

We also examined the effect of diagonal phase noise using a GKSL master
equation. For a probe-shift-motivated common-mode dephasing channel,
modeled by a projector onto the prepared state, the selected qudit Ramsey fringes
can remain visible over a broader range of dephasing strengths than the
qubit signal under the same scalar-readout convention. This robustness is
not universal for all diagonal noise. For linear Zeeman noise (\(L=J_z\)), by contrast, the higher-dimensional readouts become increasingly sensitive with dimension rather than protected. The noise resilience is therefore tied
to the symmetry of the dominant physical dephasing channel.

Overall, these results establish WM qudits, and especially qutrits, as a
realistic route to enhanced Ramsey spectroscopy and sensing at fixed
interrogation time. Future work should optimize the scalar readout,
explore composite and hyper-Ramsey versions of the WM sequence, and adapt
the protocol to the dominant noise mechanisms of specific multilevel
platforms such as trapped ions, neutral atoms, molecules, and engineered
solid-state spin systems.


\acknowledgements

This research is supported by the Bulgarian national plan for recovery and resilience, Contract No. BG-RRP-2.004-0008-C01 (SUMMIT), Project No. 3.1.4, and by the European Union’s Horizon Europe research and innovation program under Grant Agreement No. 101046968 (BRISQ).

\textit{Data availability.}---The data and the numerical code that generated the figures and Table~\ref{table:1} are available from the corresponding author upon reasonable request.

\textit{Competing interests.}---The authors declare no competing interests.


\appendix

\section{Ramsey readout metrics}
\label{app:ramsey-metrics}

This appendix defines the operational quantities used to compare Ramsey protocols of different Hilbert-space dimensions. The goal is to compare all protocols through an experimentally accessible scalar population signal.

For a general $D$-level system, let $U_D(\Delta)$ denote the full Ramsey propagator, including both the driven pulses and the free-evolution interval. If the system is initially prepared in state $|\ell\rangle$, the probability to measure it in state $|k\rangle$ is
\begin{equation}
P_{\ell\rightarrow k}(\Delta)
=
\left|
\langle k|U_D(\Delta)|\ell\rangle
\right|^2 .
\end{equation}
The experimentally useful Ramsey signal may be a single population channel or a binned readout constructed from several output channels. We therefore define the scalar readout
\begin{equation}
P_D(\Delta)
=
\sum_{k\in I_D} P_{\ell\rightarrow k}(\Delta),
\label{eq:PD}
\end{equation}
where $I_D$ is the set of measured output states included in the readout. Equivalently,
\begin{equation}
P_D(\Delta)
=
\mathrm{Tr}
\left[
\Pi_D U_D(\Delta)\rho_0 U_D^\dagger(\Delta)
\right],
\end{equation}
with $\rho_0 = |\ell\rangle\langle \ell|$, and $\Pi_D = \sum_{k\in I_D}|k\rangle\langle k|$.

We characterize the central Ramsey feature around $\Delta=0$. Let $\Delta_0$ denote the central extremum of $P_D(\Delta)$, which in the symmetric cases considered here occurs at $\Delta_0=0$. The central-fringe interval is defined as the interval between the two nearest neighboring extrema on either side of $\Delta_0$. We denote these neighboring extrema by $\Delta_-$ and $\Delta_+$, with
\[
\Delta_- < \Delta_0 < \Delta_+ .
\]
They are defined by
\begin{equation}
\left.
\frac{dP_D}{d\Delta}
\right|_{\Delta=\Delta_\pm}
=
0.
\end{equation}
The interval $[\Delta_-,\Delta_+]$ defines the central-fringe interval. All quantities $C_D$, $S_D$, and $W_D$ are extracted within this interval.

The maximum and minimum values of the scalar readout inside this interval are
$P_D^{\max} = \max_{\Delta\in[\Delta_-,\Delta_+]} P_D(\Delta)$
and
$P_D^{\min} = \min_{\Delta\in[\Delta_-,\Delta_+]} P_D(\Delta)$.
The central-fringe contrast is defined as
\begin{equation}
C_D
=
P_D^{\max}-P_D^{\min}.
\label{eq:CD}
\end{equation}
The contrast quantifies the usable amplitude of the Ramsey signal. A high-slope feature with very low contrast is not necessarily useful experimentally, because projection noise and technical noise can dominate the readout.

The maximal local slope of the central feature is defined as
\begin{equation}
S_D
=
\max_{\Delta\in[\Delta_-,\Delta_+]}
\left|
\frac{dP_D}{d\Delta}
\right|.
\label{eq:SD}
\end{equation}
The slope measures the local response of the population signal to detuning. Larger $S_D$ means that a smaller change in $\Delta$ produces a larger measurable change in the readout.

We define the midpoint of the central feature by
\begin{equation}
P_D^{\mathrm{mid}}
=
\frac{P_D^{\max}+P_D^{\min}}{2}.
\end{equation}
Let $\Delta_L$ and $\Delta_R$ be the two detunings closest to the central extremum for which
$P_D(\Delta_L) = P_D^{\mathrm{mid}}$ and $P_D(\Delta_R) = P_D^{\mathrm{mid}}$.
The full width at half contrast is then
\begin{equation}
W_D
=
|\Delta_R-\Delta_L|.
\end{equation}
A narrower feature corresponds to a smaller $W_D$. We define a dimensionless resolution-enhancement factor by comparing to the corresponding qubit width,
\begin{equation}
\mathcal{R}_D
=
\frac{W_2}{W_D}.
\end{equation}
Here $W_2$ is evaluated for the same interrogation time $\tau$ and the same convention for defining the central feature.

The quantities $S_D$, $C_D$, and $W_D$ describe the deterministic shape of the Ramsey signal. To include projection noise, we treat the scalar readout $P_D(\Delta)$ as a two-outcome measurement: the system is either detected inside the set $I_D$ or outside it. The corresponding classical Fisher information is
\begin{equation}
F_{\mathrm{cl},D}(\Delta)
=
\frac{
\left(\partial_\Delta P_D\right)^2
}{
P_D(\Delta)\left[1-P_D(\Delta)\right]
}.
\end{equation}
At points where both numerator and denominator vanish, such as the extrema of an ideal Ramsey fringe, $F_{\mathrm{cl},D}$ is understood in the limiting sense. For $N$ repetitions of the experiment, the Cramer--Rao bound~\cite{Helstrom1976,Holevo2011} gives
\begin{equation}
\delta\Delta_D(\Delta)
\geq
\frac{1}{\sqrt{N F_{\mathrm{cl},D}(\Delta)}}.
\end{equation}
Equivalently, the local projection-noise-limited uncertainty is
\begin{equation}
\delta\Delta_D(\Delta)
=
\frac{
\sqrt{P_D(\Delta)\left[1-P_D(\Delta)\right]}
}{
\sqrt{N}\left|\partial_\Delta P_D\right|}
.
\end{equation}
The optimal single-readout sensitivity within the central-fringe interval is
\begin{equation}
\delta\Delta_D^{\mathrm{opt}}
=
\min_{\Delta\in[\Delta_-,\Delta_+]}
\delta\Delta_D(\Delta)
=
\frac{1}{\sqrt{N F_{\mathrm{cl},D}^{\max}}},
\end{equation}
where
$F_{\mathrm{cl},D}^{\max}
=
\max_{\Delta\in[\Delta_-,\Delta_+]}
F_{\mathrm{cl},D}(\Delta)$.

For the ideal qubit Ramsey signal
\begin{equation}
P_2(\Delta)
=
\cos^2\left(\frac{\Delta\tau}{2}\right),
\end{equation}
we find
\begin{equation}
C_2=1,\quad
S_2=\frac{\tau}{2},\quad
W_2=\frac{\pi}{\tau},\quad
F_{\mathrm{cl},2}=\tau^2.
\end{equation}
Thus the projection-noise-limited qubit Ramsey uncertainty is
\begin{equation}
\delta\Delta_2^{\mathrm{opt}}
=
\frac{1}{\tau\sqrt{N}}.
\end{equation}

These expressions define the qubit baseline. A qudit Ramsey protocol provides a sharper detuning response when, at fixed interrogation time $\tau$, it increases $S_D$, decreases $W_D$, or improves $\delta\Delta_D^{\mathrm{opt}}$, while maintaining sufficient contrast $C_D$.

\section{Qudit Wigner--Majorana (WM) decomposition}
Here we focus on quantum systems whose dynamics exhibit SU(2) symmetry within an SU(D) representation, given by a WM decomposition~\cite{Majorana1932,stanchev2023}
\begin{align}
  \mathbf{H}_D = &\sum_{d=1}^{D} \mathbf{H}_{dd}\,|d\rangle\langle d| \nonumber \\
  &+ \sum_{d=1}^{D-1}\!\Big(\mathbf{H}_{d,d+1}\,|d\rangle\langle d{+}1|
  + \mathbf{H}_{d+1,d}\,|d{+}1\rangle\langle d| \Big),
  \label{eq:wm_hamiltonian}
\end{align}
with diagonal elements:
\begin{align}
  \mathbf{H}_{dd} = \left(d-\frac{D+1}{2}\right)\Delta,\qquad d=1,2,\ldots,D;
  \label{eq:dd_el}
\end{align}
and off-diagonal elements:
\begin{align}
  \mathbf{H}_{d+1,d} = \mathbf{H}_{d,d+1}^{\ast} = \tfrac{1}{2}\sqrt{d(D-d)}\,\Omega,\ d=1,2,\ldots,D-1.
  \label{eq:offdd_el}
\end{align}
In the magnetic-sublevel basis, the irreducible WM Hamiltonian has only nearest-neighbor ladder couplings. The Majorana construction represents the spin-$j$ propagator as the fully symmetric representation generated from an underlying spin-$1/2$ Cayley--Klein propagator. The auxiliary spin-$1/2$ amplitudes should be understood as a representation-theoretic device, not necessarily as physical qubits or physical spin-spin interactions. While Eqs.~(\ref{eq:wm_hamiltonian})-(\ref{eq:offdd_el}) exhibit the native WM spin-ladder form with nearest-neighbor couplings $\propto\sqrt{d(D\!-\!d)}\,\Omega$ \cite{Hioe1987,Varshalovich1988}, the same single SU(2) block also arises in $\Lambda$ systems under standard Raman conditions~\cite{MorrisShore1983,Zlatanov2020MS} illustrated in Fig.~\ref{fig:states}. Equations~(\ref{eq:wm_hamiltonian})--(\ref{eq:offdd_el}) are the irreducible spin-$j$ representation of SU(2), with
\begin{equation}
  D = 2j+1,\qquad m=-j,-j+1,\ldots,+j.
\end{equation}
The matrix components in this formalism are shown below. Free evolution is given by
\begin{align}
  F_D(\tau) = \sum_{d=1}^{D} e^{-i H_{dd} \tau } \, |d\rangle\langle d| .
  \label{eq:freevo}
\end{align}
with $\tau$ the interrogation time and $H_{dd}$ given by Eq.~(\ref{eq:dd_el}).

The WM Ramsey propagator used in the population-transfer plots is
\begin{equation}
U_D^{\mathrm{WM}}(\Delta)
=
R_D(T,\Delta)F_D(\tau)R_D(T,\Delta),
\end{equation}
where
\begin{equation}
R_D(T,\Delta)
=
\exp[-i\mathbf{H}_D(\Delta)T],
\qquad
\Omega T=\frac{\pi}{2}.
\end{equation}
This is the repeated-pulse convention used for the WM sequences in the main text.

We introduce the Majorana formula derived by Schwinger~\cite{Schwinger1977} to construct an arbitrary spin $j$ qudit matrix with SU(2) dynamics. 
An arbitrary angular momentum $j$ is represented by the fully symmetric sector of $2j$ spin-$1/2$ systems. Each matrix component $C_{m,m'}(t)$ of the composite spin system is given by
  \begin{align}
      C_{m,m'}&=\frac{1}{(2j)!}\sqrt{\frac{(j+m)!}{(j-m)!}\frac{(j+m')!}{(j-m')!}}\left(\frac{\partial}{\partial\alpha}\right)^{j-m}\left(\frac{\partial}{\partial\beta}\right)^{j-m'} \notag\\
      &\times\left[C_{gg}+\alpha C_{eg}+\beta C_{ge}+\alpha\beta C_{ee}\right]^{2j}
    \label{eq:Amplitude-spin}
  \end{align}
in which it is understood that $\alpha$ and $\beta$ are to be placed equal to zero after performing
the indicated differentiations.
The relations between spin $1/2$ matrix components are:
\begin{equation}
  \begin{split}
    \left|C_{gg}\right|^{2}+&\left|C_{ge}\right|^{2}=1,\\
    C^{*}_{gg}=C_{ee}&; \hspace{0.25cm} C_{eg}=-C^{*}_{ge}.
  \end{split}
  \label{eq:relations}
\end{equation}
We obtain a few examples of $(2j+1)\times(2j+1)$ complex matrices, with $C_{gg}=a$ and $C_{ge}=b$:
\begin{widetext}
\setlength{\arraycolsep}{2.5pt}
  {\scriptsize
    \begin{equation}
      \begin{split}
        &\textup{qubit},\ j = 1/2\\
        &\left(
          \begin{array}{ccc}
            a & b \\
            -b^{*} & a^{*}\\
          \end{array}
        \right),
      \end{split}
  \end{equation}}
  {\scriptsize
    \begin{equation}
      \begin{split}
        &\textup{qutrit},\ j = 1\\
        &\left(
          \begin{array}{ccc}
            a^{2} & \sqrt{2}ab & b^{2}\\
            -\sqrt{2}ab^{*} & |a|^{2}-|b|^{2} & \sqrt{2}a^{*}b\\
            b^{*2}&-\sqrt{2}a^{*}b^{*} & a^{*2}
          \end{array}
        \right),
      \end{split}
  \end{equation}}
  {\scriptsize
    \begin{equation}
      \begin{split}
        &\textup{$D=4$ qudit},\ j = 3/2\\
        &\left(
          \begin{array}{cccc}
            a^{3} & \sqrt{3}a^{2}b & \sqrt{3}ab^{2} & b^{3}\\
            -\sqrt{3}a^{2}b^{*} & a\left(|a|^{2}-2|b|^{2}\right) & b\left(2|a|^{2}-|b|^{2}\right) & \sqrt{3}a^{*}b^{2}\\
            \sqrt{3}ab^{*2} & b^{*}\left(-2|a|^{2}+|b|^{2}\right) & a^{*}\left(|a|^{2}-2|b|^{2}\right) & \sqrt{3}a^{*2}b\\
            -b^{*3} & \sqrt{3}a^{*}b^{*2} & -\sqrt{3}a^{*2}b^{} & a^{*3}\\
          \end{array}
        \right),
      \end{split}
  \end{equation}}
  {\scriptsize
    \begin{equation}
      \begin{split}
        &\textup{$D=5$},\ j = 2\\
        &\left(
          \begin{array}{ccccc}
            a^{4} & 2a^{3}b & \sqrt{6}a^{2}b^{2} & 2ab^{3} & b^{4} \\
            -2a^{3}b^{*} & a^{2}\left(|a|^{2}-3|b|^{2}\right) & \sqrt{6}ab\left(|a|^{2}-|b|^{2}\right) & b^{2}\left(3|a|^{2}-|b|^{2}\right) & 2a^{*}b^{3} \\
            \sqrt{6}a^{2}b^{*2} & \sqrt{6}ab^{*}\left(-|a|^{2}+|b|^{2}\right) & |a|^{4}-4|a|^{2}|b|^{2}+|b|^{4} & \sqrt{6}a^{*}b\left(|a|^{2}-|b|^{2}\right) & \sqrt{6}a^{*2}b^{2} \\
            -2ab^{*3} & b^{*2}\left(3|a|^{2}-|b|^{2}\right) & \sqrt{6}a^{*}b^{*}\left(-|a|^{2}+|b|^{2}\right) & a^{*2}\left(|a|^{2}-3|b|^{2}\right) & 2a^{*3}b \\
            b^{*4} & -2a^{*}b^{*3} & \sqrt{6}a^{*2}b^{*2} & -2a^{*3}b^{*} & a^{*4}
          \end{array}
        \right).
      \end{split}
  \end{equation}}
\end{widetext}
Ramsey and hyper-Ramsey matrix elements can be found in~\cite{ZanonWillette2022} to realize SU(2) high-resolution spectroscopy of a qudit.

\begin{figure}[tb]
  \centerline{\includegraphics[width=1\columnwidth]{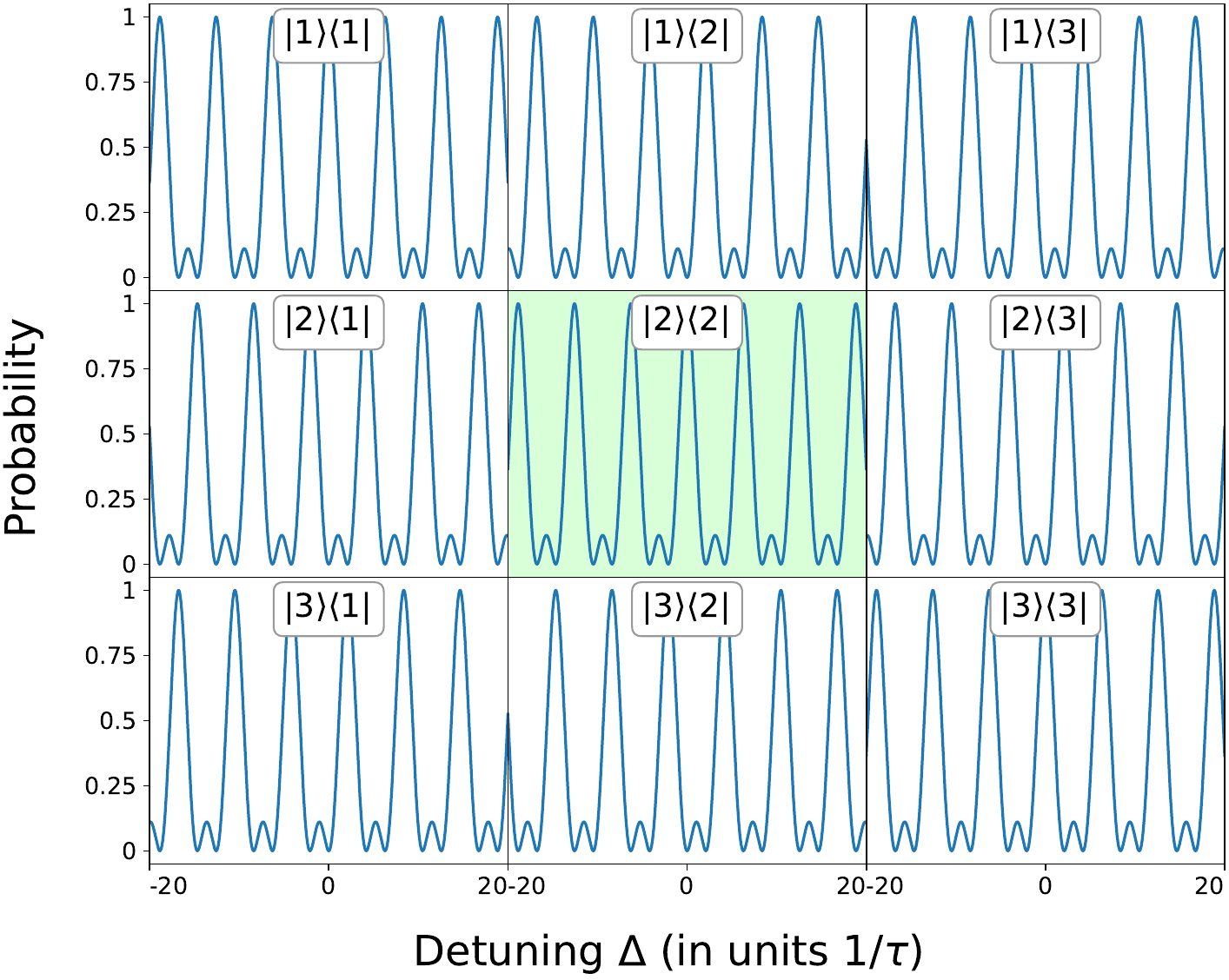}}
    \caption{Squared moduli of the propagator elements for a qutrit (three state system) versus detuning $\Delta$ for the QFT-based Ramsey sequence used in the corresponding main-text comparison. The green-highlighted panel marks the measured population-transfer channel used for the scalar QFT readout.
    }
  \label{fig:qutrit_qft_propagators}
\end{figure}

\begin{figure}[tb]
  \centerline{\includegraphics[width=1\columnwidth]{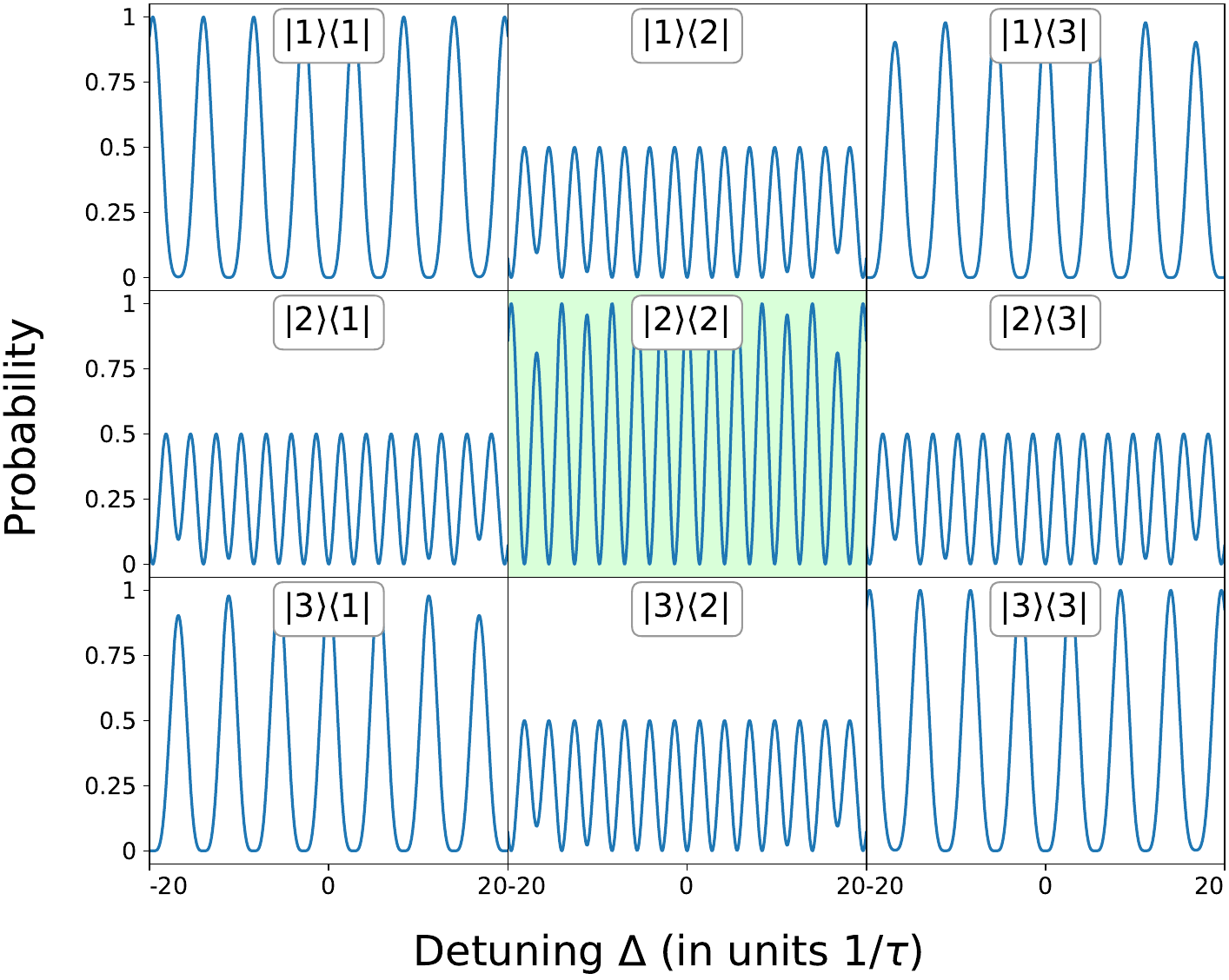}}
  \caption{
  Squared moduli of the propagator elements of the qutrit $(D=3)$ system plotted as functions of detuning $\Delta$. 
  The frame highlighted in green shows the central return probability $P_{2\rightarrow 2}(\Delta)=|\langle 2|U_3|2\rangle|^2$, which produces the oscillations presented in the main text. 
  Other parameters are identical to those used in Fig.~\ref{fig:qft}.
  }
  \label{fig:qutrit_propagators}
\end{figure}

\begin{figure}[tb]
  \centerline{\includegraphics[width=1\columnwidth]{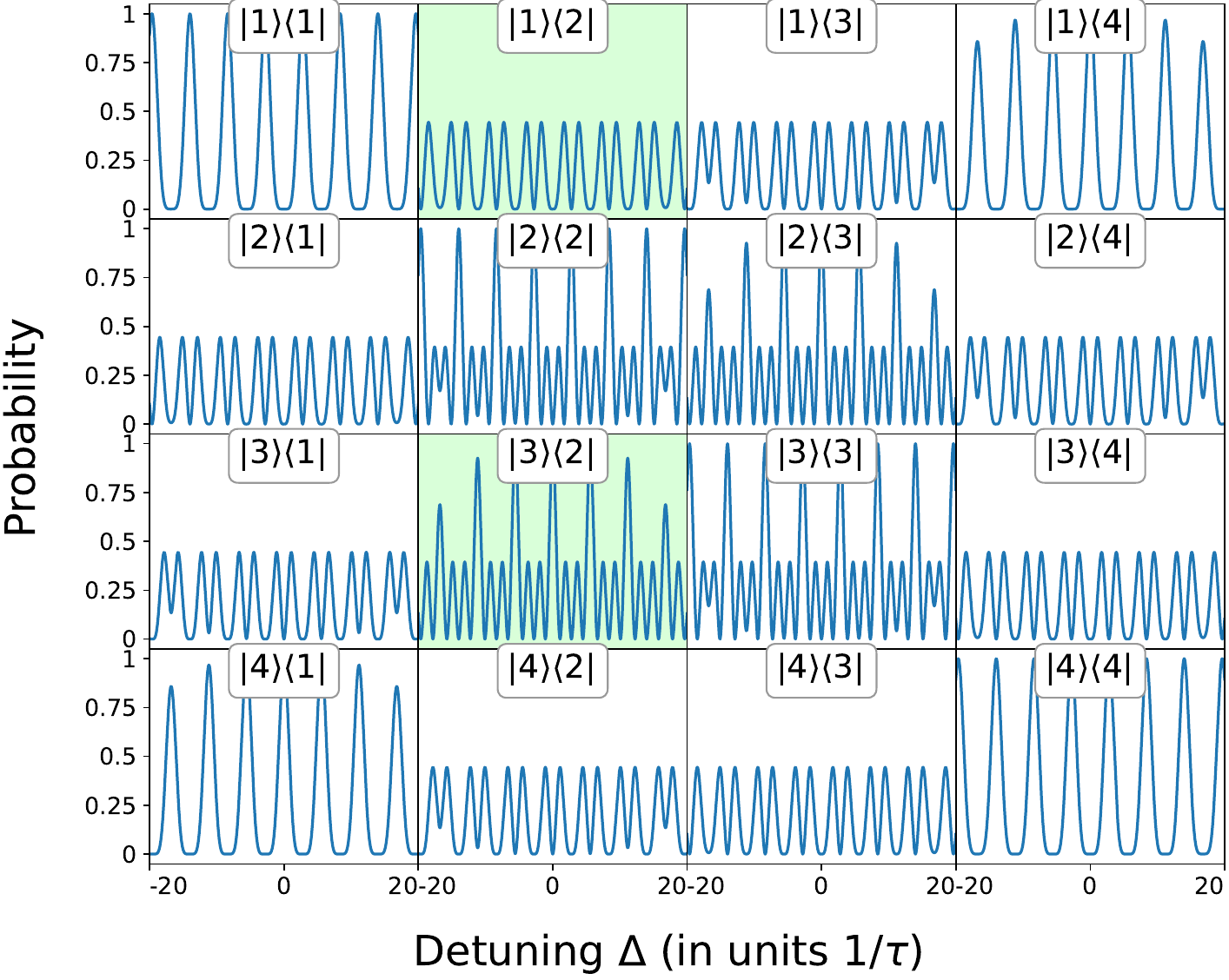}}
  \caption{
  Population-transfer elements for a $D=4$ WM qudit. 
  The green-highlighted panels show the two population-transfer elements $P_{2\rightarrow 1}(\Delta)=|\langle 1|U_4|2\rangle|^2$ and $P_{2\rightarrow 3}(\Delta)=|\langle 3|U_4|2\rangle|^2$. These are the two shoulder channels of the Ramsey interferometer for the prepared $m=-1/2$ state, and they are summed to form the scalar readout used in the main text. 
  Additionally, the propagators $P_{3\rightarrow 2}(\Delta)$ and $P_{3\rightarrow 4}(\Delta)$ can also be used to give identical oscillations for the $m=1/2$ state. 
  Other parameters are identical to those used in Fig.~\ref{fig:qft}.
  }
  \label{fig:D4_qudit_propagators}
\end{figure}

\section{Population-transfer elements of the Ramsey propagator}
\label{appendix_a}


In the main text we compare different Ramsey protocols through a scalar population readout $P_D(\Delta)$. 
This scalar signal is either a single output population, as in the qutrit case, or a binned sum of several output populations, as in the higher-dimensional WM protocols. 
The purpose of this appendix is to show the underlying population-transfer elements from which these scalar readouts are constructed.

\begin{figure}[tb]
  \centerline{\includegraphics[width=1\columnwidth]{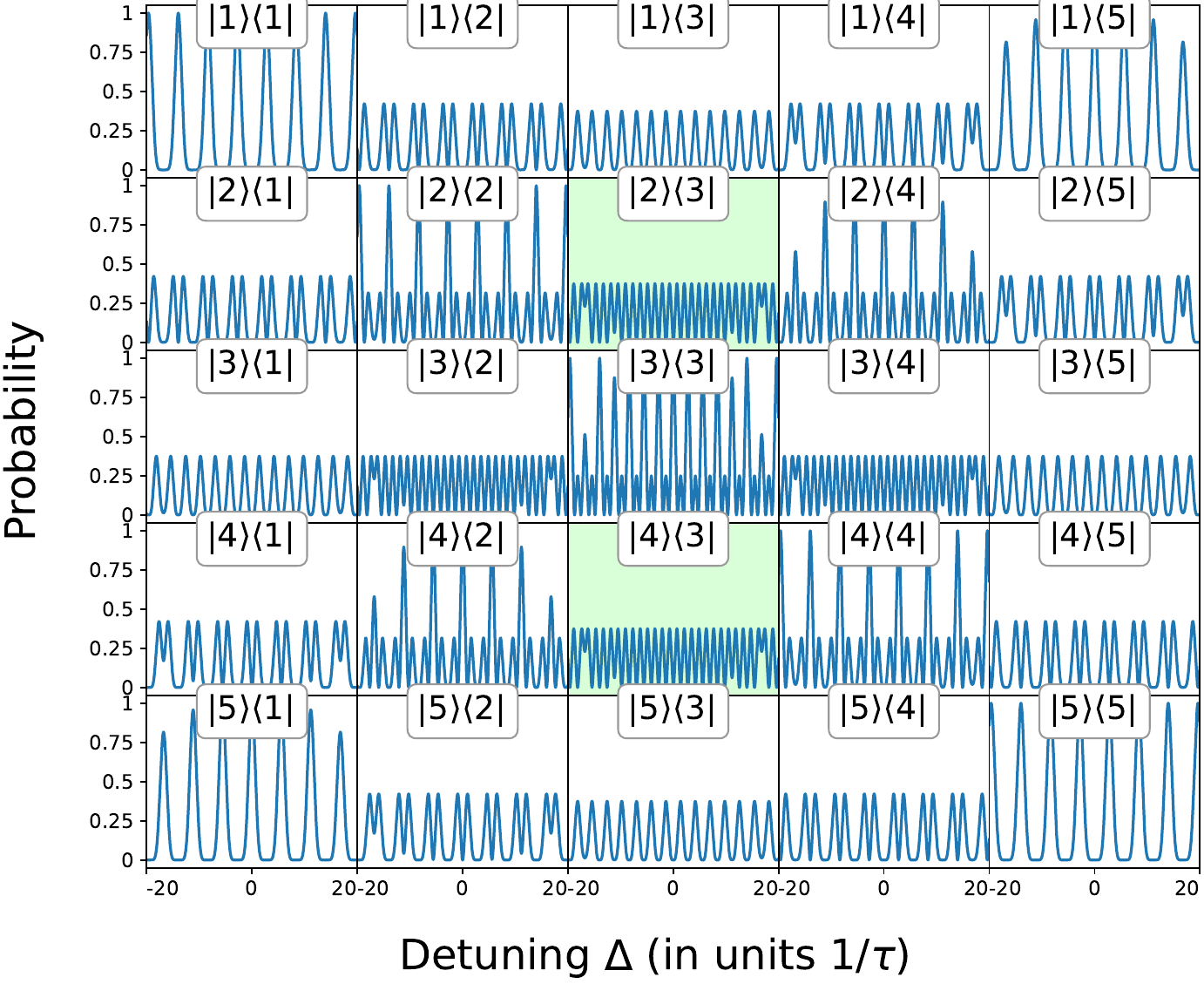}}
  \caption{Same format as Fig.~\ref{fig:D4_qudit_propagators}, but for a $D=5$ WM qudit.
  The green-highlighted panels show the two population-transfer elements $P_{3\rightarrow 2}(\Delta)=|\langle 2|U_5|3\rangle|^2$ and $P_{3\rightarrow 4}(\Delta)=|\langle 4|U_5|3\rangle|^2$. These are the two shoulder channels of the Ramsey interferometer for the prepared $m=0$ state, and they are summed to form the scalar readout used in the main text. 
  Other parameters are identical to those used in Fig.~\ref{fig:qft}.
  }
  \label{fig:D5_qudit_propagators}
\end{figure}

For a Ramsey propagator $U_D(\Delta)$, we define the population-transfer element
\begin{equation}
P_{\ell\rightarrow k}(\Delta)
=
\left|\langle k|U_D(\Delta)|\ell\rangle\right|^2 .
\end{equation}
Here $|\ell\rangle$ is the initially prepared state and $|k\rangle$ is the detected output state. 
Each panel in Figs.~\ref{fig:qutrit_qft_propagators}-\ref{fig:D5_qudit_propagators} shows one such matrix element as a function of detuning. 
The row index gives the output state $|k\rangle$, while the column index gives the input state $|\ell\rangle$. 
The highlighted panels indicate the population channels used to build the scalar Ramsey signal plotted in the main text (summarized in Table~\ref{table:readouts}).

This representation is useful for two reasons. 
First, it shows that the enhancement discussed in the paper is not a generic consequence of using a larger Hilbert space. 
The relevant information is contained in specific propagator elements, and not every multilevel superposition recombines into a useful high-contrast, high-slope scalar signal. 
Second, it explains why different readouts are used for different dimensions. For the WM qutrit, the central return probability alone captures the dominant enhanced Ramsey oscillation. For higher-dimensional WM manifolds, the near-central response is distributed mainly over two nearest-neighbor shoulder channels, so the experimentally useful readout is a binned probability.


Figure~\ref{fig:qutrit_qft_propagators} shows the population-transfer elements for the ideal qutrit QFT-based Ramsey protocol. Although the QFT creates an equal superposition over the three basis states, the highlighted measured channel does not display a clean doubling of the central Ramsey-fringe density. This supports the conclusion of Sec.~\ref{sec:qft}: abstract qudit gates such as $\mathrm{QFT}_D$ or $\sqrt{X_D}$ do not by themselves guarantee an enhanced Ramsey response. In the ideal-gate model, the gates are detuning-independent and the detuning dependence is accumulated only during the dark evolution. The resulting signal therefore depends on which coherences are created and how they are recombined by the final gate.


Figure~\ref{fig:qutrit_propagators} shows the corresponding population-transfer elements for the WM qutrit. In this case the system is prepared in the central state $|\ell\rangle=|2\rangle$ and the same state is measured at the output. The scalar readout is therefore
\begin{equation}
P_3(\Delta)
=
P_{2\rightarrow 2}(\Delta).
\end{equation}
The highlighted element $P_{2\rightarrow 2}(\Delta)$ carries the doubled central-fringe structure discussed in the main text. Physically, the first WM pulse maps the central spin-1 state onto a superposition of the two outer states. During the dark time these two components acquire opposite phases, so their relative phase evolves as $2\Delta\tau$. Recombination by the second WM pulse converts this relative phase into the central-state return probability, giving the ideal-pulse dependence
\begin{equation}
P_3(\Delta)
=
\cos^2(\Delta\tau).
\end{equation}
This is the origin of the factor-of-two compression relative to the qubit signal $\cos^2(\Delta\tau/2)$.

Figure~\ref{fig:D4_qudit_propagators} illustrates the first even-dimensional case, $D=4$. Since an even-dimensional WM manifold has no unique central state, the two central states correspond to $m=\pm 1/2$. In the convention used in the main text, the lower central state is prepared, so that $\ell=D/2=2$. The dominant near-central oscillatory response is not contained in a single return probability. Instead, it appears in the two nearest-neighbor shoulder channels adjacent to the prepared state. The scalar readout is therefore
\begin{equation}
P_4(\Delta)
=
P_{2\rightarrow 1}(\Delta)
+
P_{2\rightarrow 3}(\Delta).
\end{equation}
These are the two highlighted panels in Fig.~\ref{fig:D4_qudit_propagators}. 
By symmetry, one could equivalently prepare the upper central state $|3\rangle$ and use the pair $P_{3\rightarrow 2}(\Delta)$ and $P_{3\rightarrow 4}(\Delta)$. The two choices give equivalent Ramsey responses.

Figure~\ref{fig:D5_qudit_propagators} shows the analogous decomposition for the first higher odd-dimensional case, $D=5$. Here the WM manifold contains a unique central state, $|\ell\rangle=|3\rangle$. As in the even-dimensional case, the useful high-slope central response is distributed mainly over the two neighboring shoulder channels rather than over the return channel alone. The readout used in the main text is
\begin{equation}
P_5(\Delta)
=
P_{3\rightarrow 2}(\Delta)
+
P_{3\rightarrow 4}(\Delta).
\end{equation}
The highlighted panels identify these two contributions. This construction generalizes directly to higher odd dimensions, where the prepared state is $\ell=(D+1)/2$ and the measured channels are $\ell-1$ and $\ell+1$.

The figures in this appendix therefore provide the channel-level justification for the scalar readouts used throughout the paper. The qutrit is special because a single central return probability captures the enhanced WM interference with nearly unit contrast. In higher dimensions, the WM interference produces sharper features, but the population response is spread over several channels. The shoulder-sum readout recovers the dominant near-central contribution in a simple and experimentally accessible way, while also making explicit the resolution-contrast trade-off observed in the main text.

\bibliographystyle{apsrev4-2}
\bibliography{refs}

\end{document}